\documentclass[a4paper,twocolumn,notitlepage,nofootinbib,longbibliography,superscriptaddress,floatfix]{revtex4-2}
\usepackage{
    adjustbox,
    algorithm,
    algpseudocode,
    amsmath,
    amssymb,
    amsthm,
    booktabs,
    braket,
    comment,
    csquotes,
    enumitem,
    graphicx,
    IEEEtrantools,
    mathtools,
    multirow,
    natbib,
    physics,
    refcount,
    relsize,
    url,
    times,
    xcolor
}

\usepackage[english]{babel}
\usepackage[normalem]{ulem}
\usepackage[caption=false]{subfig}

\definecolor{mainblue}{HTML}{1f77b4}
\definecolor{mainorange}{HTML}{ff7f0e}
\definecolor{maingreen}{HTML}{2ca02c}
\definecolor{mainred}{HTML}{DC3522}
\definecolor{mainpurple}{HTML}{9467bd}
\definecolor{mainpink}{HTML}{e377c2}

\usepackage{hyperref}
\hypersetup{colorlinks=true, urlcolor=mainorange, linkcolor=maingreen, citecolor=mainblue}

\newtheorem{innercustomgeneric}{\customgenericname}
\providecommand{\customgenericname}{}
\newcommand{\newcustomtheorem}[2]{%
  \newenvironment{#1}[1]
  {%
   \renewcommand\customgenericname{#2}%
   \renewcommand\theinnercustomgeneric{##1}%
   \innercustomgeneric
  }
  {\endinnercustomgeneric}
}

\newcustomtheorem{customtheorem}{Theorem}
\newcustomtheorem{customlemma}{Lemma}
\newcustomtheorem{customproposition}{Proposition}

\begin{document}

\title{Detecting underdetermination in parameterized quantum circuits}

\author{Marie Kempkes}
\affiliation{Volkswagen Group Innovation, Berliner Ring 2, 38440 Wolfsburg, Germany}
\affiliation{Leiden University, Niels Bohrweg 1, 2333 CA Leiden, Netherlands}
\author{Jakob Spiegelberg}
\affiliation{Volkswagen Group Innovation, Berliner Ring 2, 38440 Wolfsburg, Germany}
\author{Evert van Nieuwenburg}
\affiliation{Leiden University, Niels Bohrweg 1, 2333 CA Leiden, Netherlands}
\author{Vedran Dunjko}
\affiliation{Leiden University, Niels Bohrweg 1, 2333 CA Leiden, Netherlands}

\begin{abstract}
A central question in machine learning is how reliable the predictions of a trained model are. Reliability includes the identification of instances for which a model is likely not to be trusted based on an analysis of the learning system itself. Such unreliability for an input may arise from the model family providing a variety of hypotheses consistent with the training data, which can vastly disagree in their predictions on that particular input point. This is called the underdetermination problem, and it is important to develop methods to detect it. With the emergence of quantum machine learning (QML) as a prospective alternative to classical methods for certain learning problems, the question arises to what extent they are subject to underdetermination and whether similar techniques as those developed for classical models can be employed for its detection. In this work, we first provide an overview of concepts from Safe AI and reliability, which in particular received little attention in QML. We then explore the use of a method based on local second-order information for the detection of underdetermination in parameterized quantum circuits through numerical experiments. We further demonstrate that the approach is robust to certain levels of shot noise. Our work contributes to the body of literature on Safe Quantum AI, which is an emerging field of growing importance.
\end{abstract}

\maketitle

\section{Introduction}\label{Introduction}
    The deployment of large language models to the public has emphasized the profound impact machine learning (ML) has across industry and science. Recent advancements in ML have increased the confidence in these models, driving their widespread adoption across various  domains. Despite growing enthusiasm for artificial intelligence, concerns are increasingly raised about the risks of deploying such technologies without adequate protective mechanisms. Safety-critical applications of ML, such as in medicine and autonomous driving, alongside the potential for harmful general intelligence, make it obvious that we need to establish effective controls and security measures for these systems.\\
    At the same time, advancements in the field of quantum computing suggest the potential for achieving a practically relevant quantum advantage. Despite current hardware limitations, such as a restricted number of qubits, noise during circuit execution, and trainability issues like barren plateaus~\cite{mcclean2018barren}, variational quantum algorithms (VQAs) remain a promising approach. Ongoing efforts indicate that VQAs could still offer valuable applications beyond what classical methods can achieve~\cite{gyurik2023exponential, molteni2024exponential}.\\
    While evidence demonstrating the advantage of quantum machine learning (QML) for real-world problems involving classical data is currently limited, it remains crucial to develop robust security measures before deploying these models in practical applications. We are in a uniquely advantageous position in QML to address safety considerations ahead of their actual deployment. As we elaborate in the next section, however, despite ongoing efforts in Safe QML, there are notable deficiencies in the existing literature, particularly concerning the so-called problem of reliability.\\
    
    In this work, we hence focus on a method for making predictions of QML models more reliable by identifying instances where the model's outputs are potentially \textit{not} trustworthy. We tackle the problem of \textit{underdetermination}, which indicates the extent to which hypotheses with similar performance on the training data agree or disagree about a prediction on a new test datum.\\
    To address the challenge of detecting underdetermination in the domain of QML, we apply an existing method from classical machine learning that is characterized by its theoretical soundness and computational efficiency. The method uses the Hessian matrix of the training loss function to define an underdetermination score approximating the variance of the predictions of a local ensemble (i.e., an ensemble of loss-minimizing hypotheses close to the optimal parameters found during training)~\cite{madras2020detecting}.
    For test instances that exhibit a high underdetermination score, our approach suggests that the corresponding predictions should be treated with caution and potentially disregarded in scenarios where safety-critical decisions are involved.\\    
    Our investigation focuses on the specific question of whether \textit{underdetermination in parameterized quantum circuits (PQCs) can be effectively identified using information based on the Hessian matrix}.
    Essentially, the question boils down to whether the loss landscape of PQCs around parameter settings found in the training is structured in a manner that allows local information to be sufficient for the detection of underdetermination. Our contributions can be summarized as follows:
    \begin{itemize}[noitemsep,topsep=0.0pt]
        \item We first give an overview of concepts from AI Safety and summarize works on Safe Quantum AI in order to contextualize our proposed method (Sec.~\ref{safeAI}).
        \item We demonstrate that underdetermination in parameterized quantum circuits can be detected effectively using local second-order information from the Hessian matrix of the training loss function (Sec.~\ref{numericalexperiments}), for both synthetic and real-world data.
        \item We further show that the proposed method to detect underdetermination is robust to moderate levels of shot noise and maintains a higher underdetermination detection quality than the comparative method in most situations (Sec.~\ref{numericalexperiments}).
    \end{itemize}

\section{Safe (Quantum) Artificial Intelligence}\label{safeAI}
    \begin{figure*}
    \includegraphics[width=1\textwidth]{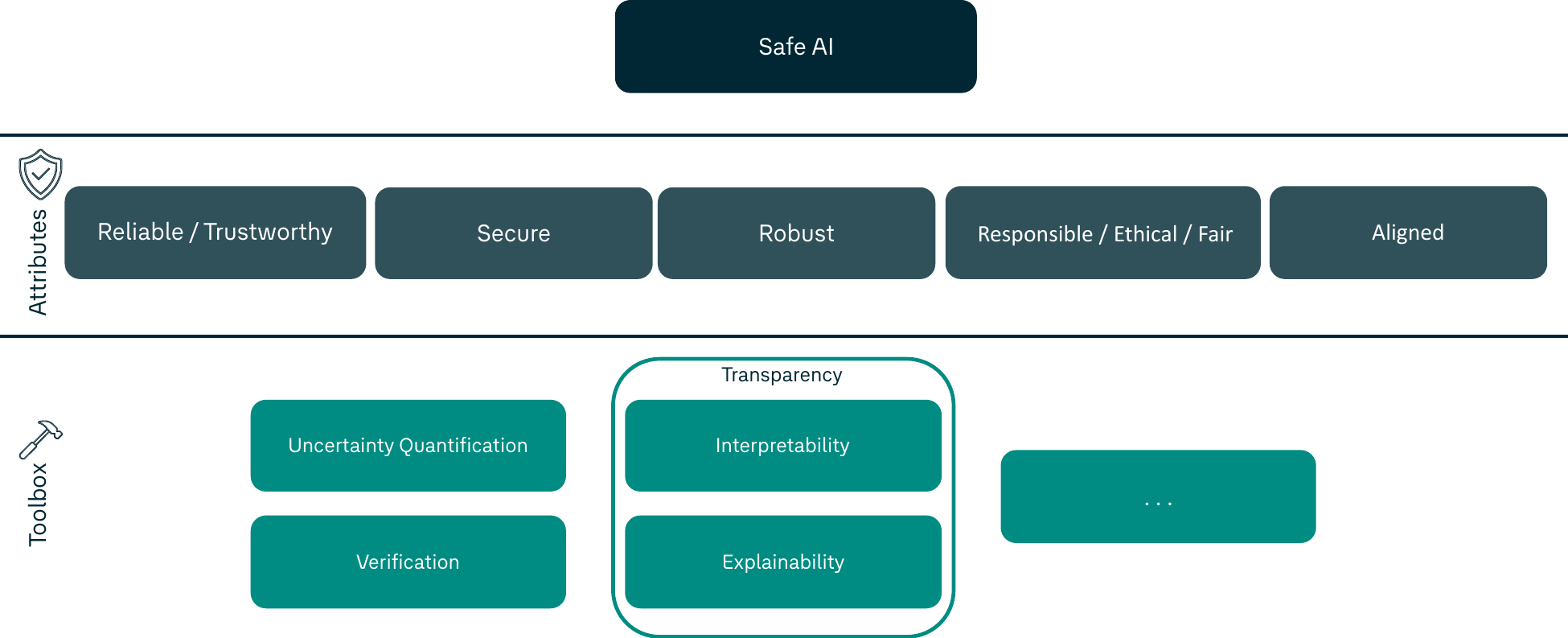}
    \caption{Overview of important concepts of Safe AI. We distinguish between attributes that are desired properties of AI systems and the toolbox that one can utilized to achieve them.}
    \centering
    \label{fig:SafeAI}
    \end{figure*}
    
    The fact that AI can pose major risks if applied incautiously is much less discussed than its capabilities, and does not impede the rapid development of new, even more powerful machine learning methods~\cite{ETO2023AIpublications}. This section outlines key concepts in Safe AI within classical machine learning and provides an overview of work in Safe Quantum AI. We note that the exact meanings of the terms we introduce here may mildly differ in literature, so the structure we provide is just one possible approach to organizing the field. The goal is to help quantum researchers identify methods that can be transferred from classical ML to QML, areas where QML can enhance classical methods, and concepts that need adaptation for the quantum domain. In Fig. \ref{fig:SafeAI} we provide an overview of important terms, which are discussed one-by-one in the following.
    
    Safe AI as an umbrella term refers to the field of research dedicated to ensuring that artificial intelligence systems are developed and deployed in a manner that minimizes potential risks for humanity. It is of great relevance especially in areas including medicine~\cite{sym13010102}, autonomous driving~\cite{muhammad2020deep}, defense~\cite{stanley2021responsible} as well as the question about long-term consequences of AI with regard to an Artificial General Intelligence (AGI)~\cite{everitt2019towards}. A selection of relevant survey papers in this field is~\cite{amodei2016concrete, mohseni2022taxonomy, juric2020ai}. A further differentiation in the usage of AI is drawn with regard to specific safety attributes that should be achieved.
    
    \textbf{Reliable / Trustworthy.}  Reliable or trustworthy AI aims at making AI systems \textit{perform as intended} across diverse environments and situations, without unexpected failures or errors. While achieving high accuracy on train and test data is important to the quality of the model, reliability adds another layer by ensuring that the model can consistently be trusted in real-world scenarios, where the data might be different from the training set~\cite{hong2023statistical}. A reliable model hence should not only make accurate predictions, but additionally provide insights into how confident it is about them. 
    
    \textbf{Secure.} In contrast, secure AI deals with safeguarding against malicious attacks, unauthorized access, and ensuring data privacy and integrity, in particular making AI invulnerable to sophisticated hacking techniques and privacy attacks~\cite{oseni2021security, cai2022generative, qayyum2021secure, liu2021privacy, liu2021whenMLmeetsprivacy, alkkassawneh2023areview}.
    
    \textbf{Robust.} AI robustness is designed to make models resilient to perturbations in the data. In contrast to security, the focus in making models robust is not on external attacks but rather on intrinsic noise due to, e.g., distributional shifts (changes in the underlying data distribution between training and inference phase)~\cite{houben2022inspect}. While a reliable model is \textit{only} required to output an appropriate confidence measure with each prediction, a robust model should remain accurate despite changing data. Taking the example of autonomous driving, suppose an autonomous driving system has been trained in America but is deployed in Europe. A \textit{reliable} model is expected to have lower confidence in its predictions in such a scenario of data drift (and, e.g., the driver has to take over steering more often). In contrast, a robust model should maintain high accuracy under such a distributional shift. Of course, achieving the latter is more challenging and guarantees on robustness are often only available for small data perturbations.

    \textbf{Responsible / Ethical / Fair.} 
    The terms responsible, ethical or fair AI, while distinct, generally refer to constructing models that are consistent with moral standards of humans, which includes behaving according to law, the inviolability of human dignity, and respecting privacy concerns. Another aspect is that AI should not exhibit spurious bias, e.g., insurance or loan decisions should not depend on ethnicity or gender~\cite{rothenberger2019relevance, mehrabi2021survey, zhang2021fairness, ryan2020ai}.
    
    \textbf{Aligned.} AI alignment deals with the question of how the training objective of AI models should be specified in order to obtain a model that matches the objective intended by the ML practitioner. An example of failed alignment would be a scenario in which the objective is to minimize the number of car-related injuries. However, the model could perfectly reach this objective by destroying all cars, which was most certainly not the intention of the practitioner. Although it partly overlaps with ideas from responsible AI, the focus in AI alignment is more on potential harm of a superintelligent machine~\cite{ji2023ai, wang2023aligning, gabriel2020artificial}.\\
    
    The aforementioned attributes can be understood as a wish list to be met by a Safe AI. This raises the question of how the individual aspects on the list can be achieved. While the complexity of data, size of the models and, ultimately, the lack of mathematical rigor in the described attributes do not allow an ultimate one-for-all solution for Safe AI, certain statements about reliability, robustness, etc. can still be made using suitable methods. For classical machine learning, numerous techniques were developed to this end, which we group together in a toolbox subdivided into different umbrella terms. It should be emphasized once again that this list is only a selection and we do not claim it to be exhaustive.
    
    \textbf{Uncertainty Quantification.} Noisy, imprecise or limited data as well as wrong model assumptions inevitably introduce uncertainties into the predictions of machine learning models. It is therefore desirable, in particular for high-stake deployments, to quantify this uncertainty so that it becomes feasible to intervene in situations of high degrees of uncertainty~\cite{huellermeier2021aleatoric, abdar2021reviewUQ}. Specifically, it was noted that standard probability distributions, such as the softmax output of a neural network, often do not capture all components of uncertainty~\cite{hullermeier2022secondorder1, hullermeier2023secondorder2}. So-called second-order predictors such as ensemble methods~\cite{lakshminarayanan2017simpleandscalable}, Bayesian neural networks~\cite{mckay1992apractical}, models based on the theory of evidence~\cite{sensoy2018evidential, li2024hyper} and conformal prediction~\cite{vovk2005algorithmic, papadopoulos2002inductive}, address this shortcoming by not only predicting probabilities for different outcomes but by simultaneously providing a distribution over these probabilities.  One aspect of uncertainty of particular importance for this work (and discussed in more detail in section \ref{method}) is underspecification and, based on this, underdetermination~\cite{damour2022underspecification, madras2020detecting}.
    
    \textbf{Verification.} Neural network verification seeks to ensure that desired properties, such as robustness to input domain perturbations, compliance with legal requirements, and adherence to fairness standards, are met. In the case of autonomous vehicles equipped with a model predicting the optimal speed of the vehicle, such a specification could be the legal speed limit, for example. An example for fairness would be that a models should not change its predictions if the only factor that has changed in the input data is the gender, which is a property that can be (approximately) formally verified. Common techniques for verification of neural networks are SMT (satisfiability modulo theory) solvers, aiming at determining whether a set of logic constraints is satisfiable~\cite{pulina2011checking, katz2019themarabou}, MIP (mixed integer programming) solvers, which optimize an objective function subject to constraints~\cite{tjeng2017evaluating, botoeva2020efficient}, as well as branch-and-bound algorithms~\cite{konig2022speeding}. We refer the interested reader to reviews covering formal verification~\cite{liu2021algorithms, meng2022adversarial, koenig2024critically}.
    
    \textbf{Interpretability.} The objective of interpretable AI is making the trained function of a neural network and hence its predictions understandable for humans. Approaches to this include, for example, comprehensible surrogate models such as (local) symbolic representations~\cite{craven1995extracting, ribeiro2016whyshould} or seeking to understand what individual layers in a neural network have learned by examining phenomena like superposition (not to be confused with superposition in quantum mechanics)~\cite{elhage2022toy}. Surveys about interpretability include~\cite{linardatos2020explainable,carvalho2019machine, zhang2021asurvey, lei2023interpretability}.
    
    \textbf{Explainability.} Explainable AI, sometimes also referred to as XAI for short, can be understood as an attenuation of interpretability: The goal here is not to fully comprehend the model, but rather to find explanations for predictions (\textit{\enquote{Why did the model decide that this image is a cat?}} or \textit{\enquote{Based on what grounds was the loan rejected?}}). Interpretability therefore always means explainability, but not vice versa. Explainability is achieved, for example, through feature importance techniques as SHAP values~\cite{vstrumbelj2014explaining}, counterfactual explanations~\cite{wachter2017counterfactual} or saliency maps~\cite{simonyan2013deep}. A selection of relevant surveys for XAI includes~\cite{minh2022explainable, saeed2023explainable, hassija2024interpreting}.\\

    Note that while one might argue that \enquote{interpretable} and \enquote{explainable} are also attributes of an AI system, they do not inherently enhance its safety. For instance, a model that provides clear explanations for its decisions may still exhibit undesirable qualities such as unreliability or unfairness.
    
    We now turn to quantum machine learning (QML) and summarize works within the field of Safe QML. A large body of literature in this domain is on security with a focus on adversarial attacks
    ~\cite{west2023towards, wendlinger2024comparative, lu2020quantumadversarial, huang2023enhancing, gong2024enhancing, liao2021robust, guan2020robustness, wiebe2018hardening, dowling2024adversarial, west2023benchmarking, winderl2024constructing, sahdev2023adversarial, huang2023certified, du2021quantumnoise} and, in particular, a review paper on secure QML was published recently~\cite{franco2024predominant}. Furthermore, research effort has been devoted to the robustness of QML models to perturbations in the input space~\cite{berberich2023training, guan2021robustness, weber2021optimal} as well as the explainability ~\cite{steinmuller2022explainable, power2024feature, liu2023quantum, heese2023explaining, shaolun2024violet, baughman2022study, gil2024opportunities} and interpretability~\cite{anschuetz2023interpretable, shaolun2023quantumeyes}. Perrier et al. further establish a foundation for fair QML~\cite{perrier2021quantumfair} discussing how it differs from its classical counterpart, while Guan et al. show how quantum noise can enhance fairness~\cite{guan2022verifying}. Additionally, Franco et al. present a hybrid quantum classical algorithm that verifies the robustness of classical neural networks and provides a polynomial speedup over classical approaches~\cite{franco2022quantum}. To the best of our knowledge, only one prior work focuses on uncertainty quantification in QML by applying a classical post-processing algorithm to obtain guarantees on the reliability of the model predictions~\cite{park2024quantumCP}. Our work complements the body of literature by introducing a reliability method based on second-order information to the quantum domain, as described in the next section.
    
\section{Underspecification, Underdetermination and Local Ensembles}\label{method}
    In this work, we aim to detect underdetermination in parameterized quantum circuits in order to enable a more reliable usage of quantum machine learning methods. This section introduces the underlying method and describes important concepts.
    
    Before considering underdetermination, we first turn to a necessary condition thereof, namely \textit{underspecification}. Underspecification describes the ambiguity of a learning algorithm for given training data and model class specification, i.e., the existence of multiple hypotheses that perform equally well on the training data. More formally, let $\mathcal{A}$ be a learning algorithm that takes training data $\mathcal{D}$ drawn from a distribution $\mathcal{P}$ as input and returns a predictor $h$ from a hypothesis class $\mathcal{H}$. We say that a learning algorithm $\mathcal{A}$ is underspecified at risk $\hat{R}$ if there exists a subset $\hat{\mathcal{H}} \subseteq \mathcal{H}$ such that for any predictor $h \in \hat{\mathcal{H}}$ returned by $\mathcal{A}$, the empirical risk $R(h) \lesssim \hat{R}$ and $|\hat{\mathcal{H}}| > 1$. 
    Different reasons for underspecification include overparameterization, noisy or unrepresentative training data and the choice of the hypothesis space.
    
    Although underspecification is not inherently problematic, it can lead to significant challenges in two distinct scenarios. First, if the training data fails to adequately represent the underlying data-generating distribution, e.g., due to a limited number of training samples, the hypotheses in $\hat{\mathcal{H}}$ may exhibit substantial disagreement when evaluated on unseen test instances drawn from the same distribution. Second, if the training data provides a representative sample of the underlying distribution, instances from the same distribution typically pose less of a concern. However, in such cases, predictions of hypotheses in $\hat{\mathcal{H}}$ can still diverge when confronted with out-of-distribution samples.
    
    A first step towards the reliable application of ML methods is therefore the ability to measure how much predictions from hypotheses in $\hat{\mathcal{H}}$ disagree on new inputs, which is referred to as the degree of \textit{underdetermination} of a prediction.
    Underdetermination thus concerns uncertainty in predictions for new inputs, whereas underspecification relates to ambiguities arising from the training data, both in conjunction with a given model.
    More formally, the degree of underdetermination for an unlabeled input $x'$ is defined by the standard deviation of predictions $\sigma(\{h\}(x'))$ of the hypotheses $h \in \hat{\mathcal{H}}$.
    Under the assumption that the given training data represents the underlying data distribution sufficiently well, the degree of underdetermination resembles a score that measures a shift in the data distribution.
    
    \begin{figure}
    \includegraphics[width=1\columnwidth]{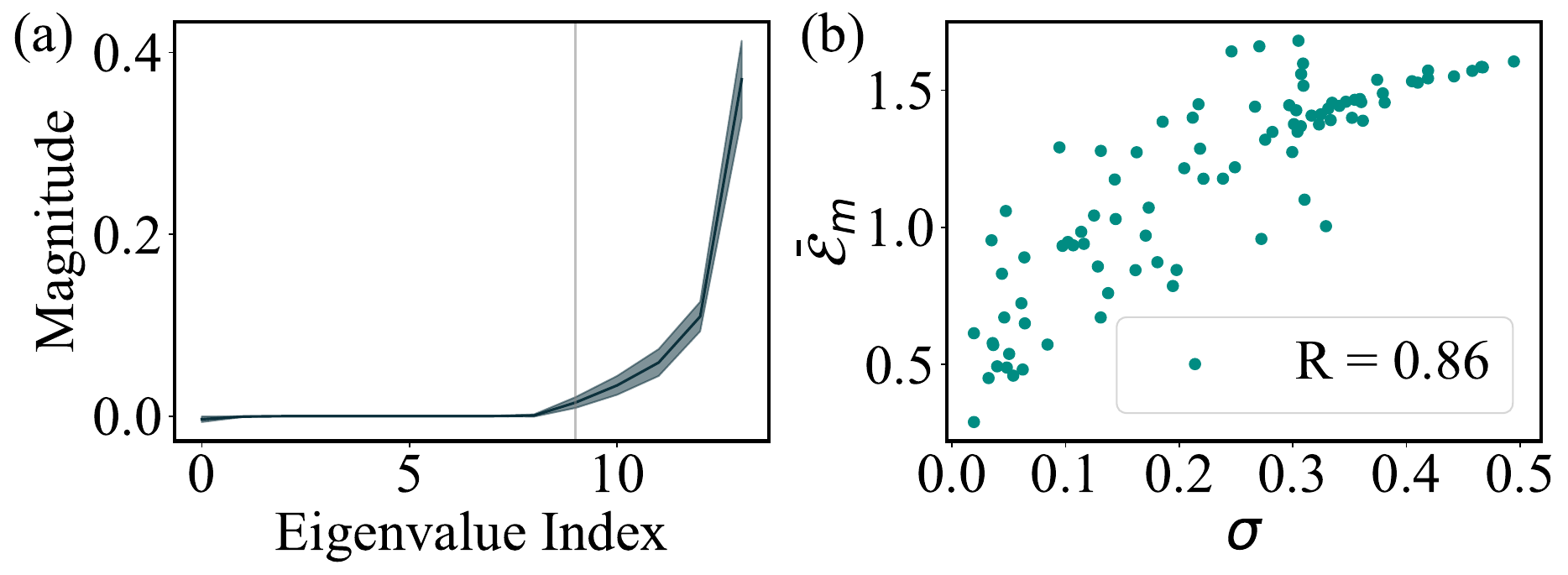}
    \caption{(a) Magnitude of eigenvalues of the Hessian matrix, thresholded at $m=6$ for the construction of $U_m$. (b) Mean extrapolation score $\bar{\mathcal{E}}_m$ is correlated to standard deviation $\sigma$ of ensemble predictions with a Pearson correlation coefficient $R=0.86$. The plot shows results for binary classification on Iris data using an ensemble of 20 PQCs.}
    \label{fig:mean_score_vs_std}
    \end{figure}

    A straightforward method for approximating underdetermination are ensemble methods, in which multiple hypotheses (e.g., obtained by varying the random seed during training) are trained on the learning task, so that the standard deviation of predictions can serve as a measure of underdetermination. We refer to this measure of underdetermination as the \textit{ensemble standard deviation}. However, these methods incur significant computational costs, not only in inference but also in training, as they require retraining with a computational overhead that scales linearly with the number of ensemble members.

    An alternative approach for approximating underdetermination that avoids aforementioned problem has been developed in~\cite{madras2020detecting}, which utilizes information based on the Hessian matrix $H$ of the cost function at the optimized parameters, defined as partial derivatives with respect to the trainable parameters $\theta$
    \begin{equation}
        (H_{\theta^*})_{\text{ij}} = \frac{\partial^2 \mathcal{L}(\theta,\mathcal{D})}{\partial \theta_i \partial \theta_j}|_{\theta=\theta^*},
    \label{eq:Hessian}
    \end{equation}
    
    evaluated at the optimized set of parameters $\theta^*$. The key concept in~\cite{madras2020detecting} is to quantify the variation of predictions of a so-called \textit{local ensemble}, which comprises hypotheses that are centered around the identified hypothesis and share comparable training costs. This measure is called the \textit{extrapolation score} $\mathcal{E}_m$ and is obtained by taking the norm of the projection of the derivative of the prediction into the subspace of low curvature, which is provably proportional to the standard deviation of a local ensemble as shown in~\cite{madras2020detecting}. 
    Intuitively, this can be understood as approximating the size of the underspecification set $\hat{H}$ locally and measuring the extent to which hypotheses within this set disagree on a new input. Since the gradient captures how outputs change with respect to parameter updates in different directions, it serves as an indicator of this disagreement.
    In the following, a detailed description of how $\mathcal{E}_m$ can be determined is given.
    
    Let $H_{\theta^*}$ be the Hessian matrix as defined in eq. (\ref{eq:Hessian}). Since it is Hermitian it can be given with the following spectral decomposition
    \begin{equation}
        H_{\theta^*} = U\,\Lambda\,U^{\dagger},
    \end{equation}
    
    where the columns of $U$ are the eigenvectors $(\xi_1,..., \xi_M)$ of $H_{\theta^*}$ and $\Lambda$ is a diagonal matrix with eigenvalues $(\lambda_1,...,\lambda_M)$ of decreasing magnitude as diagonal elements. The subspace of low curvature of the loss landscape is given by the span of eigenvectors of the Hessian matrix corresponding to small eigenvalues. Therefore, a matrix $U_m$ is defined, consisting of eigenvectors of the $(M-m)$ smallest eigenvectors of the Hessian matrix as columns, where $m$ is a hyperparameter which has to be chosen so that the subspace is \textit{sufficiently flat}. The extrapolation score is then defined as
    \begin{equation}
        \mathcal{E}_m(x^\prime) = \lVert  U_\text{m}^\dagger\, g_{\theta^*}(x^\prime) \rVert_{\text{2}},
    \label{eq:extrapolation_score}
    \end{equation}
    
    where $g_{\theta^*}(x^\prime) = \nabla_\theta \hat{y}(x^\prime, \theta^*)$ is the derivative of the prediction with respect to the parameters.
    
    The success of the extrapolation score depends heavily on the choice of the hyperparameter $m$. If $m$ is set too small, $g_{\theta^*}(x^\prime)$ is projected into well-determined regions of the loss landscape, which can render the score overly sensitive. In other words, unseen data would be attributed an overly large underdetermination score. An excessively large $m$, in contrast, can result in an insufficiently sensitive extrapolation score, which could cause underdetermined test data to not be recognized. The identification of a sound $m$ in turn depends on the eigenvalue spectrum of the Hessian matrix. A suitable $m$ can be specified for spectra showing a clear distinction between small and large eigenvalues.
    \begin{figure}
    \centering
        \includegraphics[width=1\columnwidth]{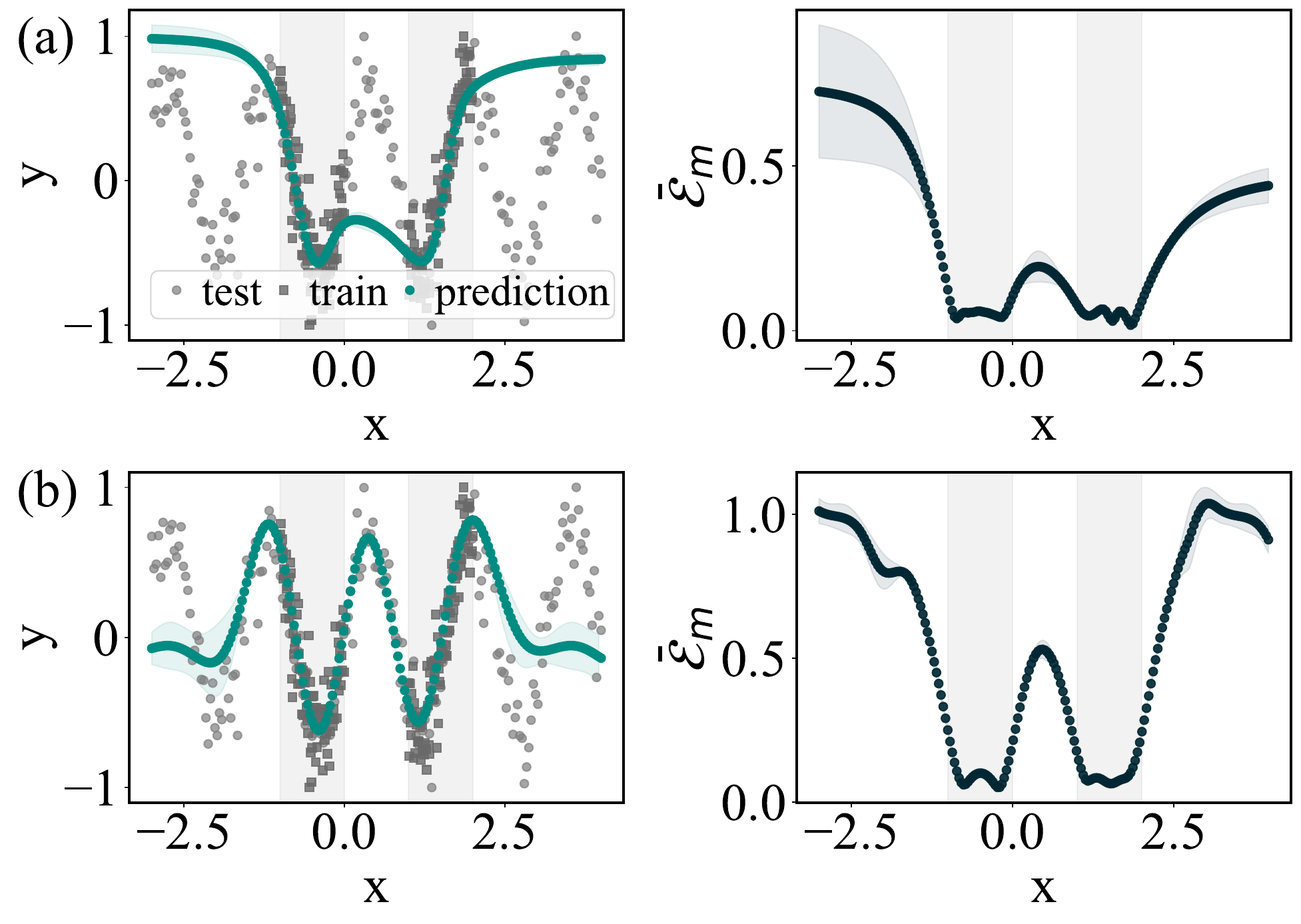}
        \caption{Predictions and mean extrapolation score $\bar{\mathcal{E}}_m$ for data sampled according to the function $y= sin(4\,x)+\mathcal{N}(0, \frac{1}{4})$ of an ensemble of 10 (a) classical neural networks and (b) parameterized quantum circuits. Training data is sampled from the grey shaded intervals only, while test data is from the full range $[-3, 4]$. The extrapolation score reliably captures underdetermination for both function families.}
    \label{fig:sinedata}
    \end{figure}

\section{Numerical Experiments}\label{numericalexperiments}
    Identifying underdetermination as proposed in the previous section is based on the idea of projecting the gradient of test predictions onto low-curvature regions of the loss landscape. With classical neural networks, it is known that the minima found in training often lie in extremely flat basins~\cite{baldassi2020shaping}, meaning that the Hessian of the loss function has many approximately zero eigenvalues at those minima. This simplifies a distinction between large and small eigenvalues, allowing for a good choice of the hyperparameter $m$ as discussed in the previous section. Less is known about the shape of the loss landscape around minima in QML models. Recent empirical studies rather indicate that the eigenvalue distribution of parameterized quantum circuits (PQCs) deviates from that of classical NNs~\cite{huembeli2021}. In this section, we therefore investigate in numerical experiments whether underdetermination in PQCs can be effectively identified using the method outlined in section~\ref{method}. We focus on PQCs as, in the same way as neural networks, they represent a parameterized function which is trained using a loss function specifying a Hessian matrix.
    \begin{figure}
    \centering
        \includegraphics[width=1\columnwidth]{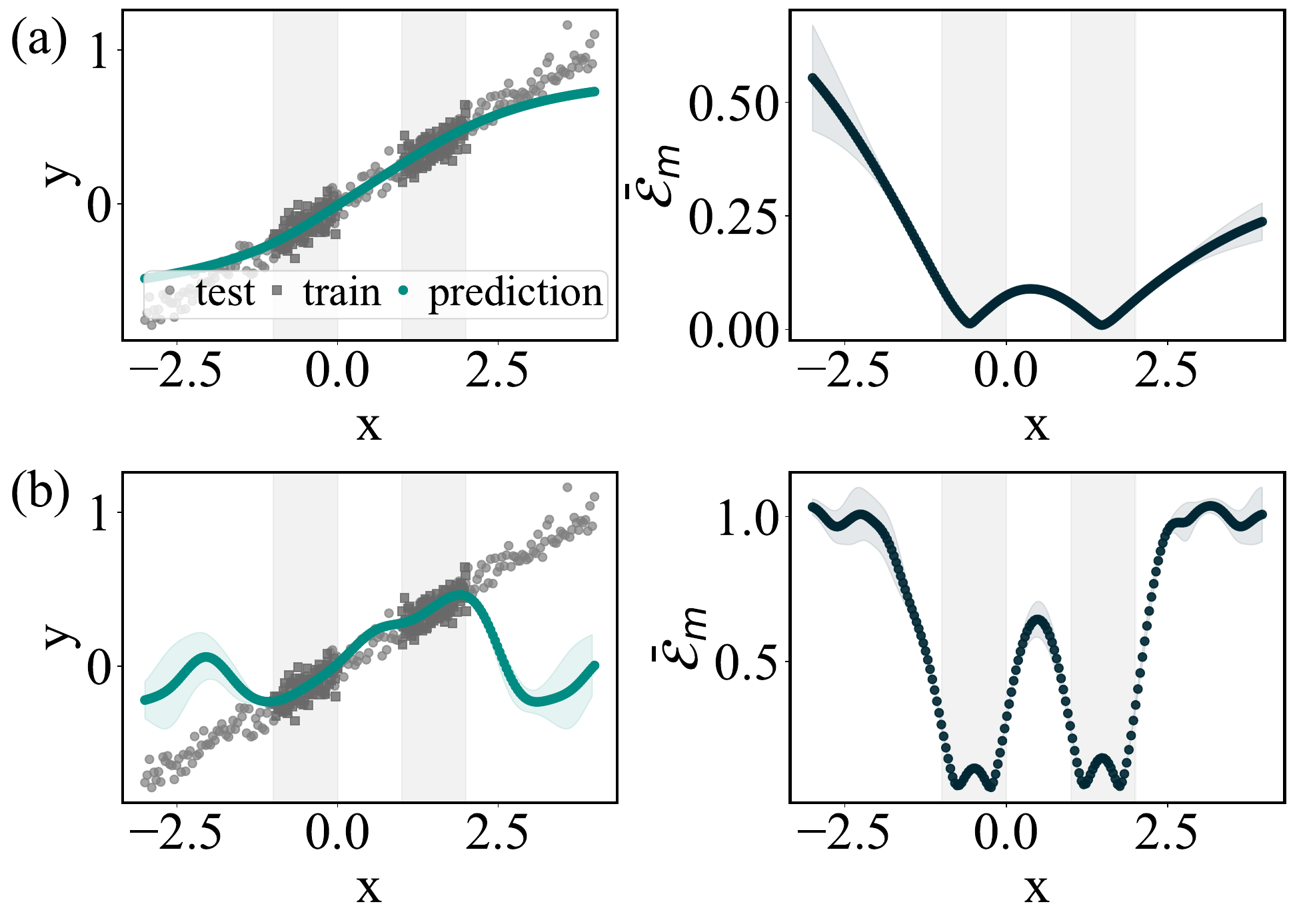}
        \caption{Predictions and mean extrapolation score $\bar{\mathcal{E}}_m$ for linear data of an ensemble of 10 (a) classical neural networks and (b) parameterized quantum circuits. Training data is sampled from the grey shaded intervals only, while test data is from the full range $[-3, 4]$. Despite significant lower predictive quality of the PQC compared to sine data predictions, the extrapolation score reliably captures underdetermination.}
    \label{fig:lineardata}
    \end{figure}
    \begin{figure*}
        \includegraphics[width=1\textwidth]{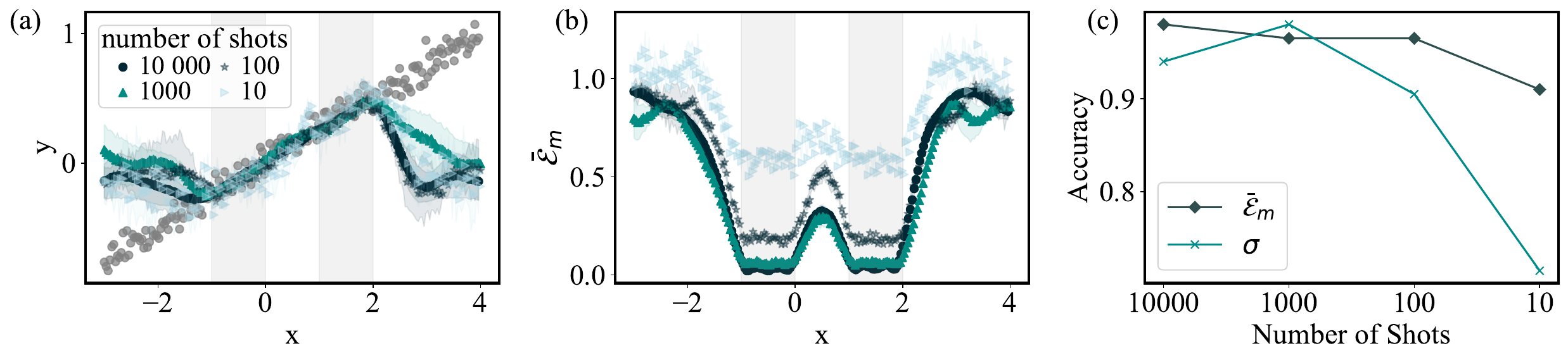}
        \caption{Investigation of the robustness of the extrapolation score $\bar{\mathcal{E}}_m$ against shot noise. Predictions (a) and extrapolation scores (b) of models with varying number of shots. In this setting, underdetermination detection boils down to classifying data inside and outside the training domain. A linear model trained on the mean extrapolation score $\bar{\mathcal{E}}_m$ achieves higher accuracy than trained on the ensemble standard deviation $\sigma$ under increasing shot noise.}
        \centering
        \label{fig:noise_analysis}
    \end{figure*}
    
    \textbf{Correlation between extrapolation score and ensemble standard deviation}.
    Given that the extrapolation score is linked to the standard deviation of a local ensemble, we first analyze the correlation between these two quantities. For the first experiment, the first two classes of the Iris dataset~\cite{iris} are considered (\textit{setosa} and \textit{versicolor}). Data is normalized to the interval $[0, \pi]$ and, as in~\cite{madras2020detecting}, split into a 10/90 train/test ratio. We train an ensemble of 20 PQCs, each of which comprises 2 qubits and 3 trainable layers. The four features of the Iris data are encoded with RZ followed by RX gates on the qubits initialized in the plus state $\ket{+}$. The trainable layer consists of RY gates on each qubit and a CNOT gate followed by RX gates on each qubit. All models are trained for 30 epochs on a batch size of 8 and attain 100\% test accuracy.
    
    The extrapolation score is determined for each ensemble member (eq. (\ref{eq:extrapolation_score})) at $m=6$ so that the eigenvectors used for constructing $U_m$ have corresponding eigenvalues sufficiently small (see Fig. \ref{fig:mean_score_vs_std} (a)). The average score is plotted against the standard deviation of the predictions, cf. Fig. \ref{fig:mean_score_vs_std} (b). We observe a Pearson correlation coefficient of $R = 0.86$, indicating a strong positive correlation. As the ensemble is not necessarily local where \enquote{local} implies that two parameter settings are not separated by regions of high loss), we do not observe perfect correlation between the two uncertainty measures. As we will see later, the ensemble standard deviation of the predictions and the extrapolation value will therefore not necessarily behave equivalently in different scenarios.
       
    \textbf{Visualization of underdetermination detection}.
    In the next experiment, we construct a scenario that allows us to assess the effectiveness of the detection of underdetermination visually. For this purpose, we closely follow~\cite{madras2020detecting} and generate one-dimensional data according to the function $y= sin(4\,x)+\mathcal{N}(0, \frac{1}{4})$, with training data only from the domain $x_\text{train}\in[-1, 0] \cap [1, 2]$ while test data is generated in the full interval $x_\text{test} \in [-3, 4]$. The used data has the property that i) underdetermination is easily visualizable (we expect high underdetermination outside the train intervals) and ii) it is low-dimensional and thus suited for small-scale quantum simulations. Ultimately, perfect underdetermination detection in this setting resembles binary classification, where data from the training intervals belongs to one class and data from outside belongs to the other.\\
    We train a PQC with 2 qubits and 3 layers, on which data is encoded on all qubits using RZ gates and each qubit is initialized in $\ket{+}$. The trainable layer consists of a RX gate on each qubit, followed by a CNOT gate and a RY gate on each qubit. Data is re-uploaded after each layer~\cite{perez2020data}. The circuit has a total of 14 trainable parameters and we choose $m = 5$ for determining the extrapolation score such that the eigenvalues are sufficiently small. In total 200 train data points and 100 test points are used, while the number of epochs is 30. In quantum machine learning, particularly with PQCs, a significant open question is how to design architectures that are both trainable and resistant to dequantization~\cite{gil2024relation, thabet2024quantum}. While theoretical considerations show that such architectures exist, identifying and constructing them remains a challenging task~\cite{liu2021rigorous, gyurik2023exponential}. Consequently, our approach focuses on leveraging PQC architectures commonly explored in the literature, without specifically considering their dequantization or trainability properties. This choice enables us to work within the current landscape of quantum models, while recognizing that the search for architectures that balance these properties is an important avenue for future research.
        
    In order to identify potential differences between parameterized quantum circuits and classical neural networks (NN), we train a neural network with the same hyperparameters as specified in~\cite{madras2020detecting}, most importantly $m=10$. 
    We show in Fig. \ref{fig:sinedata} the mean of test predictions as well as the mean extrapolation score for 10 different runs of both classical NN (a) and PQC (b). The predictions of the PQC are considerably more accurate than those of the NN, including domains in which the model has not seen any data during training, which can be explained by the inductive bias resembling the function to be learned, as PQCs can be represented via generalized trigonometric polynomials~\cite{schuld2021effectofdataencoding}.
    As shown in Fig. \ref{fig:sinedata} (right), the extrapolation score in these examples seems to be a reliable indicator of underdetermination. This becomes evident because a clear distinction could be drawn between training data and underdetermined test data outside the training domain by specifying a threshold value for the extrapolation score.
    \begin{figure}
        \centering
        \includegraphics[width=0.75\columnwidth]{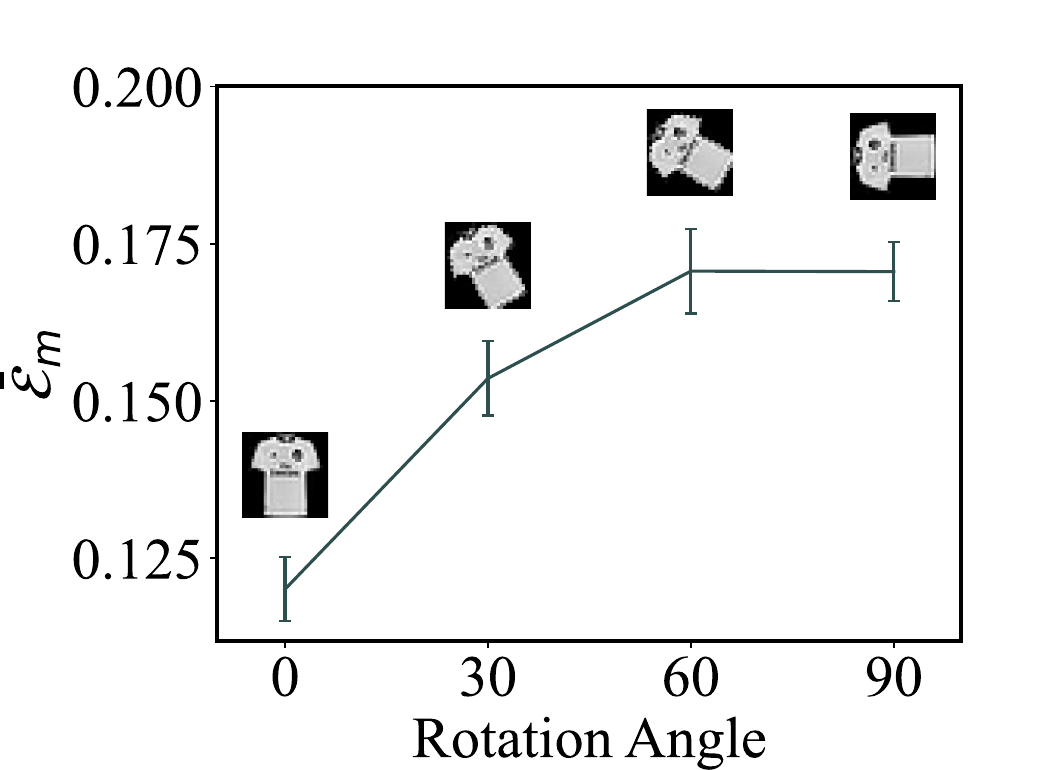}
        \caption{Mean extrapolation score increases for varying rotation angles of test images, reflecting the growing degree of distributional shift.}
        \label{fig:fmnist_angles}
    \end{figure}  
    To further verify that the reliability of the extrapolation score for the PQC is not due to the inductive bias, a different data set is investigated where a degraded performance of the PQC is to be expected, i.e., $y= \frac{1}{4} x + \mathcal{N}(0, \frac{1}{16})$. Test predictions for the same settings as before and mean extrapolation score are shown in Fig. \ref{fig:lineardata}. It becomes evident that, while the predictions of the PQC outside the training domain being significantly deteriorated compared to the classical NN, the extrapolation score remains a reliable indicator of underdetermination. Reliability in this context refers to the ability of the score to distinguish between data within the training intervals and data outside them.\\

    \textbf{Noise robustness of the extrapolation score}.
    Outputs of PQCs are defined as expectation values; however, in practical implementations, these values are estimated through sampling, which inherently introduces statistical fluctuations known as \textit{shot noise}. To estimate this, we introduce shot noise in the training stage and during inference (Fig. \ref{fig:noise_analysis} (a)). We average scores and predictions over 10 runs with different random parameter initialization. As shown in Fig. \ref{fig:noise_analysis} (b), the mean extrapolation score is reasonably robust under the tested noise strengths. It is particularly worth noting that the extrapolation score even indicates increased noise levels by displaying larger scores in the training domain. This implies that the noise causes gradients of the training inputs to shift into low curvature areas of the loss landscape, which can be detected by the extrapolation score.
    
    Lastly, we compare the robustness of the extrapolation score to that of the ensemble standard deviation $\sigma$ under shot noise. As discussed earlier, although the extrapolation score correlates with the standard deviation of a local ensemble, the question remains whether in applications a (not necessarily local) ensemble has similar properties. In our constructed learning problem, perfect underdetermination detection can be reduced to a binary classification problem (i.e., classifying whether a data point is inside or outside the training domain).
    We therefore investigate to what extent good classification is obtained using the mean extrapolation value or the standard deviation of the ensemble predictions. For this, we train a very simple Support Vector Machine (SVM) with the known labels and record the achieved accuracy for different noise levels (Fig. \ref{fig:noise_analysis} (c)). We use the \textit{sklearn} implementation of an SVM with linear kernel and default settings. It becomes evident that the mean extrapolation value provides higher accuracy in most cases, which indicates a greater robustness.
    
    \textbf{Tests on real-world data}.
    To further evaluate our method, we apply it to real-world data and larger model architectures. Specifically, we utilize the Fashion-MNIST dataset and a parameterized quantum circuit with 7 qubits and 5 trainable layers, where the data is preprocessed by applying PCA with as many dimensions as qubits. The dataset is accessible via scikit-learn~\cite{pedregosa2011scikit}. We choose the cut-off hyperparameter to be $m = 25$.
    To assess whether the extrapolation score serves as a reliable indicator of out-of-distribution samples in this setting, we train the model on two distinct classes and compute the mean score of test images with the same labels, but rotated by a specified angle. This simulates a distributional shift since the model did not see any rotated images during the training stage. Notably, the mean extrapolation score increases with larger rotation angles, reflecting the growing degree of distribution shift (Fig.~\ref{fig:fmnist_angles}).

\section{Conclusion}
    Reliable predictions are an important building block for the safe application of machine learning. In the field of quantum machine learning, however, the literature on methods for reliability is scarce. In this paper, we provide an overview of important concepts in the field of safe AI to the quantum community and further analyze a method to identify unreliable predictions. We show in numerical experiments on synthetic as well as real-world data that a score based on local second-order information is sufficient to quantify underdetermination, which is a source of uncertainty in ML. We moreover investigate the consistency of the score under shot noise and analyze its level of robustness.
    Our work therefore marks an important step toward the reliable use of quantum machine learning, paving the way for its application in safety-critical domains.

\section*{Code availability}
The code is made available from the authors upon request.

\section*{Disclaimer}
The results, opinions and conclusions expressed in this publication are not necessarily those of Volkswagen Aktiengesellschaft.

\section*{Acknowledgments}
This work was supported by the Dutch National Growth Fund (NGF), as part of the Quantum Delta NL programme as well as the Dutch Research Council (NWO/OCW), as part of the Quantum Software Consortium programme (project number 024.003.03), and co-funded by the European Union (ERC CoG, BeMAIQuantum, 101124342). Views and opinions expressed are however those of the author(s) only and do not necessarily reflect those of the European Union or the European Research Council. Neither the European Union nor the granting authority can be held responsible for them.

\bibliography{references}

\begin{thebibliography}{99}%
\makeatletter
\providecommand \@ifxundefined [1]{%
 \@ifx{#1\undefined}
}%
\providecommand \@ifnum [1]{%
 \ifnum #1\expandafter \@firstoftwo
 \else \expandafter \@secondoftwo
 \fi
}%
\providecommand \@ifx [1]{%
 \ifx #1\expandafter \@firstoftwo
 \else \expandafter \@secondoftwo
 \fi
}%
\providecommand \natexlab [1]{#1}%
\providecommand \enquote  [1]{``#1''}%
\providecommand \bibnamefont  [1]{#1}%
\providecommand \bibfnamefont [1]{#1}%
\providecommand \citenamefont [1]{#1}%
\providecommand \href@noop [0]{\@secondoftwo}%
\providecommand \href [0]{\begingroup \@sanitize@url \@href}%
\providecommand \@href[1]{\@@startlink{#1}\@@href}%
\providecommand \@@href[1]{\endgroup#1\@@endlink}%
\providecommand \@sanitize@url [0]{\catcode `\\12\catcode `\$12\catcode `\&12\catcode `\#12\catcode `\^12\catcode `\_12\catcode `\%12\relax}%
\providecommand \@@startlink[1]{}%
\providecommand \@@endlink[0]{}%
\providecommand \url  [0]{\begingroup\@sanitize@url \@url }%
\providecommand \@url [1]{\endgroup\@href {#1}{\urlprefix }}%
\providecommand \urlprefix  [0]{URL }%
\providecommand \Eprint [0]{\href }%
\providecommand \doibase [0]{https://doi.org/}%
\providecommand \selectlanguage [0]{\@gobble}%
\providecommand \bibinfo  [0]{\@secondoftwo}%
\providecommand \bibfield  [0]{\@secondoftwo}%
\providecommand \translation [1]{[#1]}%
\providecommand \BibitemOpen [0]{}%
\providecommand \bibitemStop [0]{}%
\providecommand \bibitemNoStop [0]{.\EOS\space}%
\providecommand \EOS [0]{\spacefactor3000\relax}%
\providecommand \BibitemShut  [1]{\csname bibitem#1\endcsname}%
\let\auto@bib@innerbib\@empty
%</preamble>
\bibitem [{\citenamefont {McClean}\ \emph {et~al.}(2018)\citenamefont {McClean}, \citenamefont {Boixo}, \citenamefont {Smelyanskiy}, \citenamefont {Babbush},\ and\ \citenamefont {Neven}}]{mcclean2018barren}%
  \BibitemOpen
  \bibfield  {author} {\bibinfo {author} {\bibfnamefont {J.~R.}\ \bibnamefont {McClean}}, \bibinfo {author} {\bibfnamefont {S.}~\bibnamefont {Boixo}}, \bibinfo {author} {\bibfnamefont {V.~N.}\ \bibnamefont {Smelyanskiy}}, \bibinfo {author} {\bibfnamefont {R.}~\bibnamefont {Babbush}},\ and\ \bibinfo {author} {\bibfnamefont {H.}~\bibnamefont {Neven}},\ }\bibfield  {title} {\bibinfo {title} {Barren plateaus in quantum neural network training landscapes},\ }\href@noop {} {\bibfield  {journal} {\bibinfo  {journal} {Nature communications}\ }\textbf {\bibinfo {volume} {9}},\ \bibinfo {pages} {4812} (\bibinfo {year} {2018})}\BibitemShut {NoStop}%
\bibitem [{\citenamefont {Gyurik}\ and\ \citenamefont {Dunjko}(2023)}]{gyurik2023exponential}%
  \BibitemOpen
  \bibfield  {author} {\bibinfo {author} {\bibfnamefont {C.}~\bibnamefont {Gyurik}}\ and\ \bibinfo {author} {\bibfnamefont {V.}~\bibnamefont {Dunjko}},\ }\bibfield  {title} {\bibinfo {title} {Exponential separations between classical and quantum learners},\ }\href@noop {} {\bibfield  {journal} {\bibinfo  {journal} {arXiv preprint arXiv:2306.16028}\ } (\bibinfo {year} {2023})}\BibitemShut {NoStop}%
\bibitem [{\citenamefont {Molteni}\ \emph {et~al.}(2024)\citenamefont {Molteni}, \citenamefont {Gyurik},\ and\ \citenamefont {Dunjko}}]{molteni2024exponential}%
  \BibitemOpen
  \bibfield  {author} {\bibinfo {author} {\bibfnamefont {R.}~\bibnamefont {Molteni}}, \bibinfo {author} {\bibfnamefont {C.}~\bibnamefont {Gyurik}},\ and\ \bibinfo {author} {\bibfnamefont {V.}~\bibnamefont {Dunjko}},\ }\bibfield  {title} {\bibinfo {title} {Exponential quantum advantages in learning quantum observables from classical data},\ }\href@noop {} {\bibfield  {journal} {\bibinfo  {journal} {arXiv preprint arXiv:2405.02027}\ } (\bibinfo {year} {2024})}\BibitemShut {NoStop}%
\bibitem [{\citenamefont {Madras}\ \emph {et~al.}(2020)\citenamefont {Madras}, \citenamefont {Atwood},\ and\ \citenamefont {D'Amour}}]{madras2020detecting}%
  \BibitemOpen
  \bibfield  {author} {\bibinfo {author} {\bibfnamefont {D.}~\bibnamefont {Madras}}, \bibinfo {author} {\bibfnamefont {J.}~\bibnamefont {Atwood}},\ and\ \bibinfo {author} {\bibfnamefont {A.}~\bibnamefont {D'Amour}},\ }\bibfield  {title} {\bibinfo {title} {Detecting extrapolation with local ensembles},\ }in\ \href {https://openreview.net/forum?id=BJl6bANtwH} {\emph {\bibinfo {booktitle} {International Conference on Learning Representations}}}\ (\bibinfo {year} {2020})\BibitemShut {NoStop}%
\bibitem [{\citenamefont {for Security}\ and\ \citenamefont {Technology}(2023)}]{ETO2023AIpublications}%
  \BibitemOpen
  \bibfield  {author} {\bibinfo {author} {\bibfnamefont {C.}~\bibnamefont {for Security}}\ and\ \bibinfo {author} {\bibfnamefont {E.}~\bibnamefont {Technology}},\ }\href@noop {} {\bibinfo {title} {Ai safety}},\ \bibinfo {howpublished} {\url{https://almanac.eto.tech/topics/ai-safety/}} (\bibinfo {year} {2023}),\ \bibinfo {note} {[Online: accessed 27-April-2024]}\BibitemShut {NoStop}%
\bibitem [{\citenamefont {Davahli}\ \emph {et~al.}(2021)\citenamefont {Davahli}, \citenamefont {Karwowski}, \citenamefont {Fiok}, \citenamefont {Wan},\ and\ \citenamefont {Parsaei}}]{sym13010102}%
  \BibitemOpen
  \bibfield  {author} {\bibinfo {author} {\bibfnamefont {M.~R.}\ \bibnamefont {Davahli}}, \bibinfo {author} {\bibfnamefont {W.}~\bibnamefont {Karwowski}}, \bibinfo {author} {\bibfnamefont {K.}~\bibnamefont {Fiok}}, \bibinfo {author} {\bibfnamefont {T.}~\bibnamefont {Wan}},\ and\ \bibinfo {author} {\bibfnamefont {H.~R.}\ \bibnamefont {Parsaei}},\ }\bibfield  {title} {\bibinfo {title} {Controlling safety of artificial intelligence-based systems in healthcare},\ }\bibfield  {journal} {\bibinfo  {journal} {Symmetry}\ }\textbf {\bibinfo {volume} {13}},\ \href {https://doi.org/10.3390/sym13010102} {10.3390/sym13010102} (\bibinfo {year} {2021})\BibitemShut {NoStop}%
\bibitem [{\citenamefont {Muhammad}\ \emph {et~al.}(2020)\citenamefont {Muhammad}, \citenamefont {Ullah}, \citenamefont {Lloret}, \citenamefont {Del~Ser},\ and\ \citenamefont {de~Albuquerque}}]{muhammad2020deep}%
  \BibitemOpen
  \bibfield  {author} {\bibinfo {author} {\bibfnamefont {K.}~\bibnamefont {Muhammad}}, \bibinfo {author} {\bibfnamefont {A.}~\bibnamefont {Ullah}}, \bibinfo {author} {\bibfnamefont {J.}~\bibnamefont {Lloret}}, \bibinfo {author} {\bibfnamefont {J.}~\bibnamefont {Del~Ser}},\ and\ \bibinfo {author} {\bibfnamefont {V.~H.~C.}\ \bibnamefont {de~Albuquerque}},\ }\bibfield  {title} {\bibinfo {title} {Deep learning for safe autonomous driving: Current challenges and future directions},\ }\href@noop {} {\bibfield  {journal} {\bibinfo  {journal} {IEEE Transactions on Intelligent Transportation Systems}\ }\textbf {\bibinfo {volume} {22}},\ \bibinfo {pages} {4316} (\bibinfo {year} {2020})}\BibitemShut {NoStop}%
\bibitem [{\citenamefont {Stanley-Lockman}(2021)}]{stanley2021responsible}%
  \BibitemOpen
  \bibfield  {author} {\bibinfo {author} {\bibfnamefont {Z.}~\bibnamefont {Stanley-Lockman}},\ }\href@noop {} {\emph {\bibinfo {title} {Responsible and Ethical Military AI}}}\ (\bibinfo  {publisher} {Centre for Security and Emerging Technology},\ \bibinfo {year} {2021})\BibitemShut {NoStop}%
\bibitem [{\citenamefont {Everitt}(2019)}]{everitt2019towards}%
  \BibitemOpen
  \bibfield  {author} {\bibinfo {author} {\bibfnamefont {T.}~\bibnamefont {Everitt}},\ }\emph {\bibinfo {title} {Towards Safe Artificial General Intelligence}},\ \href@noop {} {Ph.D. thesis},\ \bibinfo  {school} {The Australian National University (Australia)} (\bibinfo {year} {2019})\BibitemShut {NoStop}%
\bibitem [{\citenamefont {Amodei}\ \emph {et~al.}(2016)\citenamefont {Amodei}, \citenamefont {Olah}, \citenamefont {Steinhardt}, \citenamefont {Christiano}, \citenamefont {Schulman},\ and\ \citenamefont {Mané}}]{amodei2016concrete}%
  \BibitemOpen
  \bibfield  {author} {\bibinfo {author} {\bibfnamefont {D.}~\bibnamefont {Amodei}}, \bibinfo {author} {\bibfnamefont {C.}~\bibnamefont {Olah}}, \bibinfo {author} {\bibfnamefont {J.}~\bibnamefont {Steinhardt}}, \bibinfo {author} {\bibfnamefont {P.}~\bibnamefont {Christiano}}, \bibinfo {author} {\bibfnamefont {J.}~\bibnamefont {Schulman}},\ and\ \bibinfo {author} {\bibfnamefont {D.}~\bibnamefont {Mané}},\ }\href@noop {} {\bibinfo {title} {Concrete problems in ai safety}} (\bibinfo {year} {2016}),\ \Eprint {https://arxiv.org/abs/1606.06565} {arXiv:1606.06565} \BibitemShut {NoStop}%
\bibitem [{\citenamefont {Mohseni}\ \emph {et~al.}(2022)\citenamefont {Mohseni}, \citenamefont {Wang}, \citenamefont {Xiao}, \citenamefont {Yu}, \citenamefont {Wang},\ and\ \citenamefont {Yadawa}}]{mohseni2022taxonomy}%
  \BibitemOpen
  \bibfield  {author} {\bibinfo {author} {\bibfnamefont {S.}~\bibnamefont {Mohseni}}, \bibinfo {author} {\bibfnamefont {H.}~\bibnamefont {Wang}}, \bibinfo {author} {\bibfnamefont {C.}~\bibnamefont {Xiao}}, \bibinfo {author} {\bibfnamefont {Z.}~\bibnamefont {Yu}}, \bibinfo {author} {\bibfnamefont {Z.}~\bibnamefont {Wang}},\ and\ \bibinfo {author} {\bibfnamefont {J.}~\bibnamefont {Yadawa}},\ }\bibfield  {title} {\bibinfo {title} {Taxonomy of machine learning safety: A survey and primer},\ }\href@noop {} {\bibfield  {journal} {\bibinfo  {journal} {ACM Computing Surveys}\ }\textbf {\bibinfo {volume} {55}},\ \bibinfo {pages} {1} (\bibinfo {year} {2022})}\BibitemShut {NoStop}%
\bibitem [{\citenamefont {Juric}\ \emph {et~al.}(2020)\citenamefont {Juric}, \citenamefont {Sandic},\ and\ \citenamefont {Brcic}}]{juric2020ai}%
  \BibitemOpen
  \bibfield  {author} {\bibinfo {author} {\bibfnamefont {M.}~\bibnamefont {Juric}}, \bibinfo {author} {\bibfnamefont {A.}~\bibnamefont {Sandic}},\ and\ \bibinfo {author} {\bibfnamefont {M.}~\bibnamefont {Brcic}},\ }\bibfield  {title} {\bibinfo {title} {Ai safety: State of the field through quantitative lens},\ }in\ \href@noop {} {\emph {\bibinfo {booktitle} {2020 43rd International Convention on Information, Communication and Electronic Technology (MIPRO)}}}\ (\bibinfo {organization} {IEEE},\ \bibinfo {year} {2020})\ pp.\ \bibinfo {pages} {1254--1259}\BibitemShut {NoStop}%
\bibitem [{\citenamefont {Hong}\ \emph {et~al.}(2023)\citenamefont {Hong}, \citenamefont {Lian}, \citenamefont {Xu}, \citenamefont {Min}, \citenamefont {Wang}, \citenamefont {Freeman},\ and\ \citenamefont {Deng}}]{hong2023statistical}%
  \BibitemOpen
  \bibfield  {author} {\bibinfo {author} {\bibfnamefont {Y.}~\bibnamefont {Hong}}, \bibinfo {author} {\bibfnamefont {J.}~\bibnamefont {Lian}}, \bibinfo {author} {\bibfnamefont {L.}~\bibnamefont {Xu}}, \bibinfo {author} {\bibfnamefont {J.}~\bibnamefont {Min}}, \bibinfo {author} {\bibfnamefont {Y.}~\bibnamefont {Wang}}, \bibinfo {author} {\bibfnamefont {L.~J.}\ \bibnamefont {Freeman}},\ and\ \bibinfo {author} {\bibfnamefont {X.}~\bibnamefont {Deng}},\ }\bibfield  {title} {\bibinfo {title} {Statistical perspectives on reliability of artificial intelligence systems},\ }\href@noop {} {\bibfield  {journal} {\bibinfo  {journal} {Quality Engineering}\ }\textbf {\bibinfo {volume} {35}},\ \bibinfo {pages} {56} (\bibinfo {year} {2023})}\BibitemShut {NoStop}%
\bibitem [{\citenamefont {Oseni}\ \emph {et~al.}(2021)\citenamefont {Oseni}, \citenamefont {Moustafa}, \citenamefont {Janicke}, \citenamefont {Liu}, \citenamefont {Tari},\ and\ \citenamefont {Vasilakos}}]{oseni2021security}%
  \BibitemOpen
  \bibfield  {author} {\bibinfo {author} {\bibfnamefont {A.}~\bibnamefont {Oseni}}, \bibinfo {author} {\bibfnamefont {N.}~\bibnamefont {Moustafa}}, \bibinfo {author} {\bibfnamefont {H.}~\bibnamefont {Janicke}}, \bibinfo {author} {\bibfnamefont {P.}~\bibnamefont {Liu}}, \bibinfo {author} {\bibfnamefont {Z.}~\bibnamefont {Tari}},\ and\ \bibinfo {author} {\bibfnamefont {A.}~\bibnamefont {Vasilakos}},\ }\href@noop {} {\bibinfo {title} {Security and privacy for artificial intelligence: Opportunities and challenges}} (\bibinfo {year} {2021}),\ \Eprint {https://arxiv.org/abs/2102.04661} {arXiv:2102.04661 [cs.CR]} \BibitemShut {NoStop}%
\bibitem [{\citenamefont {Cai}\ \emph {et~al.}(2021)\citenamefont {Cai}, \citenamefont {Xiong}, \citenamefont {Xu}, \citenamefont {Wang}, \citenamefont {Li},\ and\ \citenamefont {Pan}}]{cai2022generative}%
  \BibitemOpen
  \bibfield  {author} {\bibinfo {author} {\bibfnamefont {Z.}~\bibnamefont {Cai}}, \bibinfo {author} {\bibfnamefont {Z.}~\bibnamefont {Xiong}}, \bibinfo {author} {\bibfnamefont {H.}~\bibnamefont {Xu}}, \bibinfo {author} {\bibfnamefont {P.}~\bibnamefont {Wang}}, \bibinfo {author} {\bibfnamefont {W.}~\bibnamefont {Li}},\ and\ \bibinfo {author} {\bibfnamefont {Y.}~\bibnamefont {Pan}},\ }\bibfield  {title} {\bibinfo {title} {Generative adversarial networks: A survey toward private and secure applications},\ }\bibfield  {journal} {\bibinfo  {journal} {ACM Comput. Surv.}\ }\textbf {\bibinfo {volume} {54}},\ \href {https://doi.org/10.1145/3459992} {10.1145/3459992} (\bibinfo {year} {2021})\BibitemShut {NoStop}%
\bibitem [{\citenamefont {Qayyum}\ \emph {et~al.}(2021)\citenamefont {Qayyum}, \citenamefont {Qadir}, \citenamefont {Bilal},\ and\ \citenamefont {Al-Fuqaha}}]{qayyum2021secure}%
  \BibitemOpen
  \bibfield  {author} {\bibinfo {author} {\bibfnamefont {A.}~\bibnamefont {Qayyum}}, \bibinfo {author} {\bibfnamefont {J.}~\bibnamefont {Qadir}}, \bibinfo {author} {\bibfnamefont {M.}~\bibnamefont {Bilal}},\ and\ \bibinfo {author} {\bibfnamefont {A.}~\bibnamefont {Al-Fuqaha}},\ }\bibfield  {title} {\bibinfo {title} {Secure and robust machine learning for healthcare: A survey},\ }\href {https://doi.org/10.1109/RBME.2020.3013489} {\bibfield  {journal} {\bibinfo  {journal} {IEEE Reviews in Biomedical Engineering}\ }\textbf {\bibinfo {volume} {14}},\ \bibinfo {pages} {156} (\bibinfo {year} {2021})}\BibitemShut {NoStop}%
\bibitem [{\citenamefont {Liu}\ \emph {et~al.}(2021{\natexlab{a}})\citenamefont {Liu}, \citenamefont {Xie}, \citenamefont {Wang}, \citenamefont {Zou}, \citenamefont {Xiong}, \citenamefont {Ying},\ and\ \citenamefont {Vasilakos}}]{liu2021privacy}%
  \BibitemOpen
  \bibfield  {author} {\bibinfo {author} {\bibfnamefont {X.}~\bibnamefont {Liu}}, \bibinfo {author} {\bibfnamefont {L.}~\bibnamefont {Xie}}, \bibinfo {author} {\bibfnamefont {Y.}~\bibnamefont {Wang}}, \bibinfo {author} {\bibfnamefont {J.}~\bibnamefont {Zou}}, \bibinfo {author} {\bibfnamefont {J.}~\bibnamefont {Xiong}}, \bibinfo {author} {\bibfnamefont {Z.}~\bibnamefont {Ying}},\ and\ \bibinfo {author} {\bibfnamefont {A.~V.}\ \bibnamefont {Vasilakos}},\ }\bibfield  {title} {\bibinfo {title} {Privacy and security issues in deep learning: A survey},\ }\href {https://doi.org/10.1109/ACCESS.2020.3045078} {\bibfield  {journal} {\bibinfo  {journal} {IEEE Access}\ }\textbf {\bibinfo {volume} {9}},\ \bibinfo {pages} {4566} (\bibinfo {year} {2021}{\natexlab{a}})}\BibitemShut {NoStop}%
\bibitem [{\citenamefont {Liu}\ \emph {et~al.}(2021{\natexlab{b}})\citenamefont {Liu}, \citenamefont {Ding}, \citenamefont {Shaham}, \citenamefont {Rahayu}, \citenamefont {Farokhi},\ and\ \citenamefont {Lin}}]{liu2021whenMLmeetsprivacy}%
  \BibitemOpen
  \bibfield  {author} {\bibinfo {author} {\bibfnamefont {B.}~\bibnamefont {Liu}}, \bibinfo {author} {\bibfnamefont {M.}~\bibnamefont {Ding}}, \bibinfo {author} {\bibfnamefont {S.}~\bibnamefont {Shaham}}, \bibinfo {author} {\bibfnamefont {W.}~\bibnamefont {Rahayu}}, \bibinfo {author} {\bibfnamefont {F.}~\bibnamefont {Farokhi}},\ and\ \bibinfo {author} {\bibfnamefont {Z.}~\bibnamefont {Lin}},\ }\bibfield  {title} {\bibinfo {title} {When machine learning meets privacy: A survey and outlook},\ }\bibfield  {journal} {\bibinfo  {journal} {ACM Comput. Surv.}\ }\textbf {\bibinfo {volume} {54}},\ \href {https://doi.org/10.1145/3436755} {10.1145/3436755} (\bibinfo {year} {2021}{\natexlab{b}})\BibitemShut {NoStop}%
\bibitem [{\citenamefont {Al-Khassawneh}(2023)}]{alkkassawneh2023areview}%
  \BibitemOpen
  \bibfield  {author} {\bibinfo {author} {\bibfnamefont {Y.~A.}\ \bibnamefont {Al-Khassawneh}},\ }\bibfield  {title} {\bibinfo {title} {A review of artificial intelligence in security and privacy: Research advances, applications, opportunities, and challenges},\ }\href {https://ejournal.kjpupi.id/index.php/ijost/article/view/9} {\bibfield  {journal} {\bibinfo  {journal} {Indonesian Journal of Science and Technology}\ }\textbf {\bibinfo {volume} {8}},\ \bibinfo {pages} {79–96} (\bibinfo {year} {2023})}\BibitemShut {NoStop}%
\bibitem [{\citenamefont {Houben}\ \emph {et~al.}(2022)\citenamefont {Houben}, \citenamefont {Abrecht}, \citenamefont {Akila}, \citenamefont {B{\"a}r}, \citenamefont {Brockherde}, \citenamefont {Feifel}, \citenamefont {Fingscheidt}, \citenamefont {Gannamaneni}, \citenamefont {Ghobadi}, \citenamefont {Hammam} \emph {et~al.}}]{houben2022inspect}%
  \BibitemOpen
  \bibfield  {author} {\bibinfo {author} {\bibfnamefont {S.}~\bibnamefont {Houben}}, \bibinfo {author} {\bibfnamefont {S.}~\bibnamefont {Abrecht}}, \bibinfo {author} {\bibfnamefont {M.}~\bibnamefont {Akila}}, \bibinfo {author} {\bibfnamefont {A.}~\bibnamefont {B{\"a}r}}, \bibinfo {author} {\bibfnamefont {F.}~\bibnamefont {Brockherde}}, \bibinfo {author} {\bibfnamefont {P.}~\bibnamefont {Feifel}}, \bibinfo {author} {\bibfnamefont {T.}~\bibnamefont {Fingscheidt}}, \bibinfo {author} {\bibfnamefont {S.~S.}\ \bibnamefont {Gannamaneni}}, \bibinfo {author} {\bibfnamefont {S.~E.}\ \bibnamefont {Ghobadi}}, \bibinfo {author} {\bibfnamefont {A.}~\bibnamefont {Hammam}}, \emph {et~al.},\ }\bibfield  {title} {\bibinfo {title} {Inspect, understand, overcome: A survey of practical methods for ai safety},\ }in\ \href@noop {} {\emph {\bibinfo {booktitle} {Deep Neural Networks and Data for Automated Driving: Robustness, Uncertainty Quantification, and Insights Towards Safety}}}\ (\bibinfo  {publisher} {Springer International
  Publishing Cham},\ \bibinfo {year} {2022})\ pp.\ \bibinfo {pages} {3--78}\BibitemShut {NoStop}%
\bibitem [{\citenamefont {Rothenberger}\ \emph {et~al.}(2019)\citenamefont {Rothenberger}, \citenamefont {Fabian},\ and\ \citenamefont {Arunov}}]{rothenberger2019relevance}%
  \BibitemOpen
  \bibfield  {author} {\bibinfo {author} {\bibfnamefont {L.}~\bibnamefont {Rothenberger}}, \bibinfo {author} {\bibfnamefont {B.}~\bibnamefont {Fabian}},\ and\ \bibinfo {author} {\bibfnamefont {E.}~\bibnamefont {Arunov}},\ }\bibfield  {title} {\bibinfo {title} {Relevance of ethical guidelines for artificial intelligence-a survey and evaluation.},\ }in\ \href@noop {} {\emph {\bibinfo {booktitle} {ECIS}}}\ (\bibinfo {year} {2019})\BibitemShut {NoStop}%
\bibitem [{\citenamefont {Mehrabi}\ \emph {et~al.}(2021)\citenamefont {Mehrabi}, \citenamefont {Morstatter}, \citenamefont {Saxena}, \citenamefont {Lerman},\ and\ \citenamefont {Galstyan}}]{mehrabi2021survey}%
  \BibitemOpen
  \bibfield  {author} {\bibinfo {author} {\bibfnamefont {N.}~\bibnamefont {Mehrabi}}, \bibinfo {author} {\bibfnamefont {F.}~\bibnamefont {Morstatter}}, \bibinfo {author} {\bibfnamefont {N.}~\bibnamefont {Saxena}}, \bibinfo {author} {\bibfnamefont {K.}~\bibnamefont {Lerman}},\ and\ \bibinfo {author} {\bibfnamefont {A.}~\bibnamefont {Galstyan}},\ }\bibfield  {title} {\bibinfo {title} {A survey on bias and fairness in machine learning},\ }\href@noop {} {\bibfield  {journal} {\bibinfo  {journal} {ACM computing surveys (CSUR)}\ }\textbf {\bibinfo {volume} {54}},\ \bibinfo {pages} {1} (\bibinfo {year} {2021})}\BibitemShut {NoStop}%
\bibitem [{\citenamefont {Zhang}\ and\ \citenamefont {Liu}(2021)}]{zhang2021fairness}%
  \BibitemOpen
  \bibfield  {author} {\bibinfo {author} {\bibfnamefont {X.}~\bibnamefont {Zhang}}\ and\ \bibinfo {author} {\bibfnamefont {M.}~\bibnamefont {Liu}},\ }\bibfield  {title} {\bibinfo {title} {Fairness in learning-based sequential decision algorithms: A survey},\ }in\ \href@noop {} {\emph {\bibinfo {booktitle} {Handbook of Reinforcement Learning and Control}}}\ (\bibinfo  {publisher} {Springer},\ \bibinfo {year} {2021})\ pp.\ \bibinfo {pages} {525--555}\BibitemShut {NoStop}%
\bibitem [{\citenamefont {Ryan}(2020)}]{ryan2020ai}%
  \BibitemOpen
  \bibfield  {author} {\bibinfo {author} {\bibfnamefont {M.}~\bibnamefont {Ryan}},\ }\bibfield  {title} {\bibinfo {title} {In ai we trust: Ethics, artificial intelligence, and reliability},\ }\href@noop {} {\bibfield  {journal} {\bibinfo  {journal} {Science and Engineering Ethics}\ }\textbf {\bibinfo {volume} {26}},\ \bibinfo {pages} {2749} (\bibinfo {year} {2020})}\BibitemShut {NoStop}%
\bibitem [{\citenamefont {Ji}\ \emph {et~al.}(2023)\citenamefont {Ji}, \citenamefont {Qiu}, \citenamefont {Chen}, \citenamefont {Zhang}, \citenamefont {Lou}, \citenamefont {Wang}, \citenamefont {Duan}, \citenamefont {He}, \citenamefont {Zhou}, \citenamefont {Zhang} \emph {et~al.}}]{ji2023ai}%
  \BibitemOpen
  \bibfield  {author} {\bibinfo {author} {\bibfnamefont {J.}~\bibnamefont {Ji}}, \bibinfo {author} {\bibfnamefont {T.}~\bibnamefont {Qiu}}, \bibinfo {author} {\bibfnamefont {B.}~\bibnamefont {Chen}}, \bibinfo {author} {\bibfnamefont {B.}~\bibnamefont {Zhang}}, \bibinfo {author} {\bibfnamefont {H.}~\bibnamefont {Lou}}, \bibinfo {author} {\bibfnamefont {K.}~\bibnamefont {Wang}}, \bibinfo {author} {\bibfnamefont {Y.}~\bibnamefont {Duan}}, \bibinfo {author} {\bibfnamefont {Z.}~\bibnamefont {He}}, \bibinfo {author} {\bibfnamefont {J.}~\bibnamefont {Zhou}}, \bibinfo {author} {\bibfnamefont {Z.}~\bibnamefont {Zhang}}, \emph {et~al.},\ }\bibfield  {title} {\bibinfo {title} {Ai alignment: A comprehensive survey},\ }\href@noop {} {\bibfield  {journal} {\bibinfo  {journal} {arXiv preprint arXiv:2310.19852}\ } (\bibinfo {year} {2023})}\BibitemShut {NoStop}%
\bibitem [{\citenamefont {Wang}\ \emph {et~al.}(2023)\citenamefont {Wang}, \citenamefont {Zhong}, \citenamefont {Li}, \citenamefont {Mi}, \citenamefont {Zeng}, \citenamefont {Huang}, \citenamefont {Shang}, \citenamefont {Jiang},\ and\ \citenamefont {Liu}}]{wang2023aligning}%
  \BibitemOpen
  \bibfield  {author} {\bibinfo {author} {\bibfnamefont {Y.}~\bibnamefont {Wang}}, \bibinfo {author} {\bibfnamefont {W.}~\bibnamefont {Zhong}}, \bibinfo {author} {\bibfnamefont {L.}~\bibnamefont {Li}}, \bibinfo {author} {\bibfnamefont {F.}~\bibnamefont {Mi}}, \bibinfo {author} {\bibfnamefont {X.}~\bibnamefont {Zeng}}, \bibinfo {author} {\bibfnamefont {W.}~\bibnamefont {Huang}}, \bibinfo {author} {\bibfnamefont {L.}~\bibnamefont {Shang}}, \bibinfo {author} {\bibfnamefont {X.}~\bibnamefont {Jiang}},\ and\ \bibinfo {author} {\bibfnamefont {Q.}~\bibnamefont {Liu}},\ }\bibfield  {title} {\bibinfo {title} {Aligning large language models with human: A survey},\ }\href@noop {} {\bibfield  {journal} {\bibinfo  {journal} {arXiv preprint arXiv:2307.12966}\ } (\bibinfo {year} {2023})}\BibitemShut {NoStop}%
\bibitem [{\citenamefont {Gabriel}(2020)}]{gabriel2020artificial}%
  \BibitemOpen
  \bibfield  {author} {\bibinfo {author} {\bibfnamefont {I.}~\bibnamefont {Gabriel}},\ }\bibfield  {title} {\bibinfo {title} {Artificial intelligence, values, and alignment},\ }\href@noop {} {\bibfield  {journal} {\bibinfo  {journal} {Minds and machines}\ }\textbf {\bibinfo {volume} {30}},\ \bibinfo {pages} {411} (\bibinfo {year} {2020})}\BibitemShut {NoStop}%
\bibitem [{\citenamefont {H{\"u}llermeier}\ and\ \citenamefont {Waegeman}(2021)}]{huellermeier2021aleatoric}%
  \BibitemOpen
  \bibfield  {author} {\bibinfo {author} {\bibfnamefont {E.}~\bibnamefont {H{\"u}llermeier}}\ and\ \bibinfo {author} {\bibfnamefont {W.}~\bibnamefont {Waegeman}},\ }\bibfield  {title} {\bibinfo {title} {Aleatoric and {E}pistemic {U}ncertainty in {M}achine {L}earning: {A}n {I}ntroduction to {C}oncepts and {M}ethods},\ }\href@noop {} {\bibfield  {journal} {\bibinfo  {journal} {Machine Learning}\ }\textbf {\bibinfo {volume} {110}},\ \bibinfo {pages} {457} (\bibinfo {year} {2021})}\BibitemShut {NoStop}%
\bibitem [{\citenamefont {Abdar}\ \emph {et~al.}(2021)\citenamefont {Abdar}, \citenamefont {Pourpanah}, \citenamefont {Hussain}, \citenamefont {Rezazadegan}, \citenamefont {Liu}, \citenamefont {Ghavamzadeh}, \citenamefont {Fieguth}, \citenamefont {Cao}, \citenamefont {Khosravi}, \citenamefont {Acharya}, \citenamefont {Makarenkov},\ and\ \citenamefont {Nahavandi}}]{abdar2021reviewUQ}%
  \BibitemOpen
  \bibfield  {author} {\bibinfo {author} {\bibfnamefont {M.}~\bibnamefont {Abdar}}, \bibinfo {author} {\bibfnamefont {F.}~\bibnamefont {Pourpanah}}, \bibinfo {author} {\bibfnamefont {S.}~\bibnamefont {Hussain}}, \bibinfo {author} {\bibfnamefont {D.}~\bibnamefont {Rezazadegan}}, \bibinfo {author} {\bibfnamefont {L.}~\bibnamefont {Liu}}, \bibinfo {author} {\bibfnamefont {M.}~\bibnamefont {Ghavamzadeh}}, \bibinfo {author} {\bibfnamefont {P.}~\bibnamefont {Fieguth}}, \bibinfo {author} {\bibfnamefont {X.}~\bibnamefont {Cao}}, \bibinfo {author} {\bibfnamefont {A.}~\bibnamefont {Khosravi}}, \bibinfo {author} {\bibfnamefont {U.~R.}\ \bibnamefont {Acharya}}, \bibinfo {author} {\bibfnamefont {V.}~\bibnamefont {Makarenkov}},\ and\ \bibinfo {author} {\bibfnamefont {S.}~\bibnamefont {Nahavandi}},\ }\bibfield  {title} {\bibinfo {title} {A review of uncertainty quantification in deep learning: Techniques, applications and challenges},\ }\href {https://doi.org/https://doi.org/10.1016/j.inffus.2021.05.008} {\bibfield
  {journal} {\bibinfo  {journal} {Information Fusion}\ }\textbf {\bibinfo {volume} {76}},\ \bibinfo {pages} {243} (\bibinfo {year} {2021})}\BibitemShut {NoStop}%
\bibitem [{\citenamefont {Bengs}\ \emph {et~al.}(2022)\citenamefont {Bengs}, \citenamefont {H{\"u}llermeier},\ and\ \citenamefont {Waegeman}}]{hullermeier2022secondorder1}%
  \BibitemOpen
  \bibfield  {author} {\bibinfo {author} {\bibfnamefont {V.}~\bibnamefont {Bengs}}, \bibinfo {author} {\bibfnamefont {E.}~\bibnamefont {H{\"u}llermeier}},\ and\ \bibinfo {author} {\bibfnamefont {W.}~\bibnamefont {Waegeman}},\ }\bibfield  {title} {\bibinfo {title} {Pitfalls of epistemic uncertainty quantification through loss minimisation},\ }in\ \href@noop {} {\emph {\bibinfo {booktitle} {Neural Information Processing Systems}}}\ (\bibinfo {year} {2022})\BibitemShut {NoStop}%
\bibitem [{\citenamefont {Bengs}\ \emph {et~al.}(2023)\citenamefont {Bengs}, \citenamefont {H\"{u}llermeier},\ and\ \citenamefont {Waegeman}}]{hullermeier2023secondorder2}%
  \BibitemOpen
  \bibfield  {author} {\bibinfo {author} {\bibfnamefont {V.}~\bibnamefont {Bengs}}, \bibinfo {author} {\bibfnamefont {E.}~\bibnamefont {H\"{u}llermeier}},\ and\ \bibinfo {author} {\bibfnamefont {W.}~\bibnamefont {Waegeman}},\ }\bibfield  {title} {\bibinfo {title} {On second-order scoring rules for epistemic uncertainty quantification},\ }in\ \href@noop {} {\emph {\bibinfo {booktitle} {Proceedings of the 40th International Conference on Machine Learning}}},\ Vol.\ \bibinfo {volume} {202},\ \bibinfo {editor} {edited by\ \bibinfo {editor} {\bibfnamefont {A.}~\bibnamefont {Krause}}, \bibinfo {editor} {\bibfnamefont {E.}~\bibnamefont {Brunskill}}, \bibinfo {editor} {\bibfnamefont {K.}~\bibnamefont {Cho}}, \bibinfo {editor} {\bibfnamefont {B.}~\bibnamefont {Engelhardt}}, \bibinfo {editor} {\bibfnamefont {S.}~\bibnamefont {Sabato}},\ and\ \bibinfo {editor} {\bibfnamefont {J.}~\bibnamefont {Scarlett}}}\ (\bibinfo  {publisher} {PMLR},\ \bibinfo {year} {2023})\ pp.\ \bibinfo {pages} {2078--2091}\BibitemShut {NoStop}%
\bibitem [{\citenamefont {Lakshminarayanan}\ \emph {et~al.}(2017)\citenamefont {Lakshminarayanan}, \citenamefont {Pritzel},\ and\ \citenamefont {Blundell}}]{lakshminarayanan2017simpleandscalable}%
  \BibitemOpen
  \bibfield  {author} {\bibinfo {author} {\bibfnamefont {B.}~\bibnamefont {Lakshminarayanan}}, \bibinfo {author} {\bibfnamefont {A.}~\bibnamefont {Pritzel}},\ and\ \bibinfo {author} {\bibfnamefont {C.}~\bibnamefont {Blundell}},\ }\bibfield  {title} {\bibinfo {title} {Simple and scalable predictive uncertainty estimation using deep ensembles},\ }in\ \href@noop {} {\emph {\bibinfo {booktitle} {Proceedings of the 31st International Conference on Neural Information Processing Systems}}}\ (\bibinfo  {publisher} {Curran Associates Inc.},\ \bibinfo {address} {Red Hook, NY, USA},\ \bibinfo {year} {2017})\ p.\ \bibinfo {pages} {6405–6416}\BibitemShut {NoStop}%
\bibitem [{\citenamefont {MacKay}(1992)}]{mckay1992apractical}%
  \BibitemOpen
  \bibfield  {author} {\bibinfo {author} {\bibfnamefont {D.~J.~C.}\ \bibnamefont {MacKay}},\ }\bibfield  {title} {\bibinfo {title} {A practical bayesian framework for backpropagation networks},\ }\href {https://doi.org/10.1162/neco.1992.4.3.448} {\bibfield  {journal} {\bibinfo  {journal} {Neural Computation}\ }\textbf {\bibinfo {volume} {4}},\ \bibinfo {pages} {448} (\bibinfo {year} {1992})}\BibitemShut {NoStop}%
\bibitem [{\citenamefont {Sensoy}\ \emph {et~al.}(2018)\citenamefont {Sensoy}, \citenamefont {Kaplan},\ and\ \citenamefont {Kandemir}}]{sensoy2018evidential}%
  \BibitemOpen
  \bibfield  {author} {\bibinfo {author} {\bibfnamefont {M.}~\bibnamefont {Sensoy}}, \bibinfo {author} {\bibfnamefont {L.}~\bibnamefont {Kaplan}},\ and\ \bibinfo {author} {\bibfnamefont {M.}~\bibnamefont {Kandemir}},\ }\bibfield  {title} {\bibinfo {title} {Evidential deep learning to quantify classification uncertainty},\ }\href@noop {} {\bibfield  {journal} {\bibinfo  {journal} {Advances in neural information processing systems}\ }\textbf {\bibinfo {volume} {31}} (\bibinfo {year} {2018})}\BibitemShut {NoStop}%
\bibitem [{\citenamefont {Li}\ \emph {et~al.}(2024)\citenamefont {Li}, \citenamefont {Li}, \citenamefont {Ou}, \citenamefont {Kaplan}, \citenamefont {J{\o}sang}, \citenamefont {Cho}, \citenamefont {JEONG},\ and\ \citenamefont {Chen}}]{li2024hyper}%
  \BibitemOpen
  \bibfield  {author} {\bibinfo {author} {\bibfnamefont {C.}~\bibnamefont {Li}}, \bibinfo {author} {\bibfnamefont {K.}~\bibnamefont {Li}}, \bibinfo {author} {\bibfnamefont {Y.}~\bibnamefont {Ou}}, \bibinfo {author} {\bibfnamefont {L.~M.}\ \bibnamefont {Kaplan}}, \bibinfo {author} {\bibfnamefont {A.}~\bibnamefont {J{\o}sang}}, \bibinfo {author} {\bibfnamefont {J.-H.}\ \bibnamefont {Cho}}, \bibinfo {author} {\bibfnamefont {D.~H.}\ \bibnamefont {JEONG}},\ and\ \bibinfo {author} {\bibfnamefont {F.}~\bibnamefont {Chen}},\ }\bibfield  {title} {\bibinfo {title} {Hyper evidential deep learning to quantify composite classification uncertainty},\ }in\ \href {https://openreview.net/forum?id=A7t7z6g6tM} {\emph {\bibinfo {booktitle} {The Twelfth International Conference on Learning Representations}}}\ (\bibinfo {year} {2024})\BibitemShut {NoStop}%
\bibitem [{\citenamefont {Vovk}\ \emph {et~al.}(2005)\citenamefont {Vovk}, \citenamefont {Gammerman},\ and\ \citenamefont {Shafer}}]{vovk2005algorithmic}%
  \BibitemOpen
  \bibfield  {author} {\bibinfo {author} {\bibfnamefont {V.}~\bibnamefont {Vovk}}, \bibinfo {author} {\bibfnamefont {A.}~\bibnamefont {Gammerman}},\ and\ \bibinfo {author} {\bibfnamefont {G.}~\bibnamefont {Shafer}},\ }\href@noop {} {\emph {\bibinfo {title} {Algorithmic Learning in a Random World}}},\ Vol.~\bibinfo {volume} {29}\ (\bibinfo  {publisher} {Springer},\ \bibinfo {year} {2005})\BibitemShut {NoStop}%
\bibitem [{\citenamefont {Papadopoulos}\ \emph {et~al.}(2002)\citenamefont {Papadopoulos}, \citenamefont {Proedrou}, \citenamefont {Vovk},\ and\ \citenamefont {Gammerman}}]{papadopoulos2002inductive}%
  \BibitemOpen
  \bibfield  {author} {\bibinfo {author} {\bibfnamefont {H.}~\bibnamefont {Papadopoulos}}, \bibinfo {author} {\bibfnamefont {K.}~\bibnamefont {Proedrou}}, \bibinfo {author} {\bibfnamefont {V.}~\bibnamefont {Vovk}},\ and\ \bibinfo {author} {\bibfnamefont {A.}~\bibnamefont {Gammerman}},\ }\bibfield  {title} {\bibinfo {title} {Inductive confidence machines for regression},\ }in\ \href@noop {} {\emph {\bibinfo {booktitle} {Machine Learning: ECML 2002}}},\ \bibinfo {editor} {edited by\ \bibinfo {editor} {\bibfnamefont {T.}~\bibnamefont {Elomaa}}, \bibinfo {editor} {\bibfnamefont {H.}~\bibnamefont {Mannila}},\ and\ \bibinfo {editor} {\bibfnamefont {H.}~\bibnamefont {Toivonen}}}\ (\bibinfo  {publisher} {Springer Berlin Heidelberg},\ \bibinfo {address} {Berlin, Heidelberg},\ \bibinfo {year} {2002})\ pp.\ \bibinfo {pages} {345--356}\BibitemShut {NoStop}%
\bibitem [{\citenamefont {D'Amour}\ \emph {et~al.}(2022)\citenamefont {D'Amour}, \citenamefont {Heller}, \citenamefont {Moldovan}, \citenamefont {Adlam}, \citenamefont {Alipanahi}, \citenamefont {Beutel}, \citenamefont {Chen}, \citenamefont {Deaton}, \citenamefont {Eisenstein}, \citenamefont {Hoffman}, \citenamefont {Hormozdiari}, \citenamefont {Houlsby}, \citenamefont {Hou}, \citenamefont {Jerfel}, \citenamefont {Karthikesalingam}, \citenamefont {Lucic}, \citenamefont {Ma}, \citenamefont {McLean}, \citenamefont {Mincu}, \citenamefont {Mitani}, \citenamefont {Montanari}, \citenamefont {Nado}, \citenamefont {Natarajan}, \citenamefont {Nielson}, \citenamefont {Osborne}, \citenamefont {Raman}, \citenamefont {Ramasamy}, \citenamefont {Sayres}, \citenamefont {Schrouff}, \citenamefont {Seneviratne}, \citenamefont {Sequeira}, \citenamefont {Suresh}, \citenamefont {Veitch}, \citenamefont {Vladymyrov}, \citenamefont {Wang}, \citenamefont {Webster}, \citenamefont {Yadlowsky}, \citenamefont {Yun}, \citenamefont {Zhai},\
  and\ \citenamefont {Sculley}}]{damour2022underspecification}%
  \BibitemOpen
  \bibfield  {author} {\bibinfo {author} {\bibfnamefont {A.}~\bibnamefont {D'Amour}}, \bibinfo {author} {\bibfnamefont {K.}~\bibnamefont {Heller}}, \bibinfo {author} {\bibfnamefont {D.}~\bibnamefont {Moldovan}}, \bibinfo {author} {\bibfnamefont {B.}~\bibnamefont {Adlam}}, \bibinfo {author} {\bibfnamefont {B.}~\bibnamefont {Alipanahi}}, \bibinfo {author} {\bibfnamefont {A.}~\bibnamefont {Beutel}}, \bibinfo {author} {\bibfnamefont {C.}~\bibnamefont {Chen}}, \bibinfo {author} {\bibfnamefont {J.}~\bibnamefont {Deaton}}, \bibinfo {author} {\bibfnamefont {J.}~\bibnamefont {Eisenstein}}, \bibinfo {author} {\bibfnamefont {M.~D.}\ \bibnamefont {Hoffman}}, \bibinfo {author} {\bibfnamefont {F.}~\bibnamefont {Hormozdiari}}, \bibinfo {author} {\bibfnamefont {N.}~\bibnamefont {Houlsby}}, \bibinfo {author} {\bibfnamefont {S.}~\bibnamefont {Hou}}, \bibinfo {author} {\bibfnamefont {G.}~\bibnamefont {Jerfel}}, \bibinfo {author} {\bibfnamefont {A.}~\bibnamefont {Karthikesalingam}}, \bibinfo {author} {\bibfnamefont
  {M.}~\bibnamefont {Lucic}}, \bibinfo {author} {\bibfnamefont {Y.}~\bibnamefont {Ma}}, \bibinfo {author} {\bibfnamefont {C.}~\bibnamefont {McLean}}, \bibinfo {author} {\bibfnamefont {D.}~\bibnamefont {Mincu}}, \bibinfo {author} {\bibfnamefont {A.}~\bibnamefont {Mitani}}, \bibinfo {author} {\bibfnamefont {A.}~\bibnamefont {Montanari}}, \bibinfo {author} {\bibfnamefont {Z.}~\bibnamefont {Nado}}, \bibinfo {author} {\bibfnamefont {V.}~\bibnamefont {Natarajan}}, \bibinfo {author} {\bibfnamefont {C.}~\bibnamefont {Nielson}}, \bibinfo {author} {\bibfnamefont {T.~F.}\ \bibnamefont {Osborne}}, \bibinfo {author} {\bibfnamefont {R.}~\bibnamefont {Raman}}, \bibinfo {author} {\bibfnamefont {K.}~\bibnamefont {Ramasamy}}, \bibinfo {author} {\bibfnamefont {R.}~\bibnamefont {Sayres}}, \bibinfo {author} {\bibfnamefont {J.}~\bibnamefont {Schrouff}}, \bibinfo {author} {\bibfnamefont {M.}~\bibnamefont {Seneviratne}}, \bibinfo {author} {\bibfnamefont {S.}~\bibnamefont {Sequeira}}, \bibinfo {author} {\bibfnamefont
  {H.}~\bibnamefont {Suresh}}, \bibinfo {author} {\bibfnamefont {V.}~\bibnamefont {Veitch}}, \bibinfo {author} {\bibfnamefont {M.}~\bibnamefont {Vladymyrov}}, \bibinfo {author} {\bibfnamefont {X.}~\bibnamefont {Wang}}, \bibinfo {author} {\bibfnamefont {K.}~\bibnamefont {Webster}}, \bibinfo {author} {\bibfnamefont {S.}~\bibnamefont {Yadlowsky}}, \bibinfo {author} {\bibfnamefont {T.}~\bibnamefont {Yun}}, \bibinfo {author} {\bibfnamefont {X.}~\bibnamefont {Zhai}},\ and\ \bibinfo {author} {\bibfnamefont {D.}~\bibnamefont {Sculley}},\ }\bibfield  {title} {\bibinfo {title} {Underspecification presents challenges for credibility in modern machine learning},\ }\href {https://dl.acm.org/doi/abs/10.5555/3586589.3586815} {\bibfield  {journal} {\bibinfo  {journal} {J. Mach. Learn. Res.}\ }\textbf {\bibinfo {volume} {23}} (\bibinfo {year} {2022})}\BibitemShut {NoStop}%
\bibitem [{\citenamefont {Pulina}\ and\ \citenamefont {Tacchella}(2011)}]{pulina2011checking}%
  \BibitemOpen
  \bibfield  {author} {\bibinfo {author} {\bibfnamefont {L.}~\bibnamefont {Pulina}}\ and\ \bibinfo {author} {\bibfnamefont {A.}~\bibnamefont {Tacchella}},\ }\bibfield  {title} {\bibinfo {title} {Checking safety of neural networks with smt solvers: A comparative evaluation},\ }in\ \href@noop {} {\emph {\bibinfo {booktitle} {Congress of the Italian Association for Artificial Intelligence}}}\ (\bibinfo {organization} {Springer},\ \bibinfo {year} {2011})\ pp.\ \bibinfo {pages} {127--138}\BibitemShut {NoStop}%
\bibitem [{\citenamefont {Katz}\ \emph {et~al.}(2019)\citenamefont {Katz}, \citenamefont {Huang}, \citenamefont {Ibeling}, \citenamefont {Julian}, \citenamefont {Lazarus}, \citenamefont {Lim}, \citenamefont {Shah}, \citenamefont {Thakoor}, \citenamefont {Wu}, \citenamefont {Zelji{\'{c}}}, \citenamefont {Dill}, \citenamefont {Kochenderfer},\ and\ \citenamefont {Barrett}}]{katz2019themarabou}%
  \BibitemOpen
  \bibfield  {author} {\bibinfo {author} {\bibfnamefont {G.}~\bibnamefont {Katz}}, \bibinfo {author} {\bibfnamefont {D.~A.}\ \bibnamefont {Huang}}, \bibinfo {author} {\bibfnamefont {D.}~\bibnamefont {Ibeling}}, \bibinfo {author} {\bibfnamefont {K.}~\bibnamefont {Julian}}, \bibinfo {author} {\bibfnamefont {C.}~\bibnamefont {Lazarus}}, \bibinfo {author} {\bibfnamefont {R.}~\bibnamefont {Lim}}, \bibinfo {author} {\bibfnamefont {P.}~\bibnamefont {Shah}}, \bibinfo {author} {\bibfnamefont {S.}~\bibnamefont {Thakoor}}, \bibinfo {author} {\bibfnamefont {H.}~\bibnamefont {Wu}}, \bibinfo {author} {\bibfnamefont {A.}~\bibnamefont {Zelji{\'{c}}}}, \bibinfo {author} {\bibfnamefont {D.~L.}\ \bibnamefont {Dill}}, \bibinfo {author} {\bibfnamefont {M.~J.}\ \bibnamefont {Kochenderfer}},\ and\ \bibinfo {author} {\bibfnamefont {C.}~\bibnamefont {Barrett}},\ }\bibfield  {title} {\bibinfo {title} {The marabou framework for verification and analysis of deep neural networks},\ }in\ \href@noop {} {\emph {\bibinfo {booktitle}
  {Computer Aided Verification}}},\ \bibinfo {editor} {edited by\ \bibinfo {editor} {\bibfnamefont {I.}~\bibnamefont {Dillig}}\ and\ \bibinfo {editor} {\bibfnamefont {S.}~\bibnamefont {Tasiran}}}\ (\bibinfo  {publisher} {Springer International Publishing},\ \bibinfo {address} {Cham},\ \bibinfo {year} {2019})\ pp.\ \bibinfo {pages} {443--452}\BibitemShut {NoStop}%
\bibitem [{\citenamefont {Tjeng}\ \emph {et~al.}(2017)\citenamefont {Tjeng}, \citenamefont {Xiao},\ and\ \citenamefont {Tedrake}}]{tjeng2017evaluating}%
  \BibitemOpen
  \bibfield  {author} {\bibinfo {author} {\bibfnamefont {V.}~\bibnamefont {Tjeng}}, \bibinfo {author} {\bibfnamefont {K.~Y.}\ \bibnamefont {Xiao}},\ and\ \bibinfo {author} {\bibfnamefont {R.}~\bibnamefont {Tedrake}},\ }\bibfield  {title} {\bibinfo {title} {Evaluating robustness of neural networks with mixed integer programming},\ }in\ \href {https://api.semanticscholar.org/CorpusID:47016770} {\emph {\bibinfo {booktitle} {International Conference on Learning Representations}}}\ (\bibinfo {year} {2017})\BibitemShut {NoStop}%
\bibitem [{\citenamefont {Botoeva}\ \emph {et~al.}(2020)\citenamefont {Botoeva}, \citenamefont {Kouvaros}, \citenamefont {Kronqvist}, \citenamefont {Lomuscio},\ and\ \citenamefont {Misener}}]{botoeva2020efficient}%
  \BibitemOpen
  \bibfield  {author} {\bibinfo {author} {\bibfnamefont {E.}~\bibnamefont {Botoeva}}, \bibinfo {author} {\bibfnamefont {P.}~\bibnamefont {Kouvaros}}, \bibinfo {author} {\bibfnamefont {J.}~\bibnamefont {Kronqvist}}, \bibinfo {author} {\bibfnamefont {A.}~\bibnamefont {Lomuscio}},\ and\ \bibinfo {author} {\bibfnamefont {R.}~\bibnamefont {Misener}},\ }\bibfield  {title} {\bibinfo {title} {Efficient verification of relu-based neural networks via dependency analysis},\ }\href {https://doi.org/10.1609/aaai.v34i04.5729} {\bibfield  {journal} {\bibinfo  {journal} {Proceedings of the AAAI Conference on Artificial Intelligence}\ }\textbf {\bibinfo {volume} {34}},\ \bibinfo {pages} {3291} (\bibinfo {year} {2020})}\BibitemShut {NoStop}%
\bibitem [{\citenamefont {K{\"o}nig}\ \emph {et~al.}(2022)\citenamefont {K{\"o}nig}, \citenamefont {Hoos},\ and\ \citenamefont {Rijn}}]{konig2022speeding}%
  \BibitemOpen
  \bibfield  {author} {\bibinfo {author} {\bibfnamefont {M.}~\bibnamefont {K{\"o}nig}}, \bibinfo {author} {\bibfnamefont {H.~H.}\ \bibnamefont {Hoos}},\ and\ \bibinfo {author} {\bibfnamefont {J.~N.~v.}\ \bibnamefont {Rijn}},\ }\bibfield  {title} {\bibinfo {title} {Speeding up neural network robustness verification via algorithm configuration and an optimised mixed integer linear programming solver portfolio},\ }\href@noop {} {\bibfield  {journal} {\bibinfo  {journal} {Machine Learning}\ }\textbf {\bibinfo {volume} {111}},\ \bibinfo {pages} {4565} (\bibinfo {year} {2022})}\BibitemShut {NoStop}%
\bibitem [{\citenamefont {Liu}\ \emph {et~al.}(2021{\natexlab{c}})\citenamefont {Liu}, \citenamefont {Arnon}, \citenamefont {Lazarus}, \citenamefont {Strong}, \citenamefont {Barrett}, \citenamefont {Kochenderfer} \emph {et~al.}}]{liu2021algorithms}%
  \BibitemOpen
  \bibfield  {author} {\bibinfo {author} {\bibfnamefont {C.}~\bibnamefont {Liu}}, \bibinfo {author} {\bibfnamefont {T.}~\bibnamefont {Arnon}}, \bibinfo {author} {\bibfnamefont {C.}~\bibnamefont {Lazarus}}, \bibinfo {author} {\bibfnamefont {C.}~\bibnamefont {Strong}}, \bibinfo {author} {\bibfnamefont {C.}~\bibnamefont {Barrett}}, \bibinfo {author} {\bibfnamefont {M.~J.}\ \bibnamefont {Kochenderfer}}, \emph {et~al.},\ }\bibfield  {title} {\bibinfo {title} {Algorithms for verifying deep neural networks},\ }\href@noop {} {\bibfield  {journal} {\bibinfo  {journal} {Foundations and Trends{\textregistered} in Optimization}\ }\textbf {\bibinfo {volume} {4}},\ \bibinfo {pages} {244} (\bibinfo {year} {2021}{\natexlab{c}})}\BibitemShut {NoStop}%
\bibitem [{\citenamefont {Meng}\ \emph {et~al.}(2022)\citenamefont {Meng}, \citenamefont {Bai}, \citenamefont {Teo}, \citenamefont {Hou}, \citenamefont {Xiao}, \citenamefont {Lin},\ and\ \citenamefont {Dong}}]{meng2022adversarial}%
  \BibitemOpen
  \bibfield  {author} {\bibinfo {author} {\bibfnamefont {M.~H.}\ \bibnamefont {Meng}}, \bibinfo {author} {\bibfnamefont {G.}~\bibnamefont {Bai}}, \bibinfo {author} {\bibfnamefont {S.~G.}\ \bibnamefont {Teo}}, \bibinfo {author} {\bibfnamefont {Z.}~\bibnamefont {Hou}}, \bibinfo {author} {\bibfnamefont {Y.}~\bibnamefont {Xiao}}, \bibinfo {author} {\bibfnamefont {Y.}~\bibnamefont {Lin}},\ and\ \bibinfo {author} {\bibfnamefont {J.~S.}\ \bibnamefont {Dong}},\ }\bibfield  {title} {\bibinfo {title} {Adversarial robustness of deep neural networks: A survey from a formal verification perspective},\ }\href {https://doi.org/10.1109/TDSC.2022.3179131} {\bibfield  {journal} {\bibinfo  {journal} {IEEE Transactions on Dependable and Secure Computing}\ ,\ \bibinfo {pages} {1}} (\bibinfo {year} {2022})}\BibitemShut {NoStop}%
\bibitem [{\citenamefont {K{{\"o}}nig}\ \emph {et~al.}(2024)\citenamefont {K{{\"o}}nig}, \citenamefont {Bosman}, \citenamefont {Hoos},\ and\ \citenamefont {van Rijn}}]{koenig2024critically}%
  \BibitemOpen
  \bibfield  {author} {\bibinfo {author} {\bibfnamefont {M.}~\bibnamefont {K{{\"o}}nig}}, \bibinfo {author} {\bibfnamefont {A.~W.}\ \bibnamefont {Bosman}}, \bibinfo {author} {\bibfnamefont {H.~H.}\ \bibnamefont {Hoos}},\ and\ \bibinfo {author} {\bibfnamefont {J.~N.}\ \bibnamefont {van Rijn}},\ }\bibfield  {title} {\bibinfo {title} {Critically assessing the state of the art in neural network verification},\ }\href {http://jmlr.org/papers/v25/23-0119.html} {\bibfield  {journal} {\bibinfo  {journal} {Journal of Machine Learning Research}\ }\textbf {\bibinfo {volume} {25}},\ \bibinfo {pages} {1} (\bibinfo {year} {2024})}\BibitemShut {NoStop}%
\bibitem [{\citenamefont {Craven}\ and\ \citenamefont {Shavlik}(1995)}]{craven1995extracting}%
  \BibitemOpen
  \bibfield  {author} {\bibinfo {author} {\bibfnamefont {M.~W.}\ \bibnamefont {Craven}}\ and\ \bibinfo {author} {\bibfnamefont {J.~W.}\ \bibnamefont {Shavlik}},\ }\bibfield  {title} {\bibinfo {title} {Extracting tree-structured representations of trained networks},\ }in\ \href@noop {} {\emph {\bibinfo {booktitle} {Proceedings of the 8th International Conference on Neural Information Processing Systems}}},\ \bibinfo {series and number} {NIPS'95}\ (\bibinfo  {publisher} {MIT Press},\ \bibinfo {address} {Cambridge, MA, USA},\ \bibinfo {year} {1995})\ p.\ \bibinfo {pages} {24–30}\BibitemShut {NoStop}%
\bibitem [{\citenamefont {Ribeiro}\ \emph {et~al.}(2016)\citenamefont {Ribeiro}, \citenamefont {Singh},\ and\ \citenamefont {Guestrin}}]{ribeiro2016whyshould}%
  \BibitemOpen
  \bibfield  {author} {\bibinfo {author} {\bibfnamefont {M.~T.}\ \bibnamefont {Ribeiro}}, \bibinfo {author} {\bibfnamefont {S.}~\bibnamefont {Singh}},\ and\ \bibinfo {author} {\bibfnamefont {C.}~\bibnamefont {Guestrin}},\ }\bibfield  {title} {\bibinfo {title} {"why should i trust you?" explaining the predictions of any classifier},\ }in\ \href@noop {} {\emph {\bibinfo {booktitle} {Proceedings of the 22nd ACM SIGKDD international conference on knowledge discovery and data mining}}}\ (\bibinfo {year} {2016})\ pp.\ \bibinfo {pages} {1135--1144}\BibitemShut {NoStop}%
\bibitem [{\citenamefont {Elhage}\ \emph {et~al.}(2022)\citenamefont {Elhage}, \citenamefont {Hume}, \citenamefont {Olsson}, \citenamefont {Schiefer}, \citenamefont {Henighan}, \citenamefont {Kravec}, \citenamefont {Hatfield-Dodds}, \citenamefont {Lasenby}, \citenamefont {Drain}, \citenamefont {Chen} \emph {et~al.}}]{elhage2022toy}%
  \BibitemOpen
  \bibfield  {author} {\bibinfo {author} {\bibfnamefont {N.}~\bibnamefont {Elhage}}, \bibinfo {author} {\bibfnamefont {T.}~\bibnamefont {Hume}}, \bibinfo {author} {\bibfnamefont {C.}~\bibnamefont {Olsson}}, \bibinfo {author} {\bibfnamefont {N.}~\bibnamefont {Schiefer}}, \bibinfo {author} {\bibfnamefont {T.}~\bibnamefont {Henighan}}, \bibinfo {author} {\bibfnamefont {S.}~\bibnamefont {Kravec}}, \bibinfo {author} {\bibfnamefont {Z.}~\bibnamefont {Hatfield-Dodds}}, \bibinfo {author} {\bibfnamefont {R.}~\bibnamefont {Lasenby}}, \bibinfo {author} {\bibfnamefont {D.}~\bibnamefont {Drain}}, \bibinfo {author} {\bibfnamefont {C.}~\bibnamefont {Chen}}, \emph {et~al.},\ }\bibfield  {title} {\bibinfo {title} {Toy models of superposition},\ }\href@noop {} {\bibfield  {journal} {\bibinfo  {journal} {arXiv preprint arXiv:2209.10652}\ } (\bibinfo {year} {2022})}\BibitemShut {NoStop}%
\bibitem [{\citenamefont {Linardatos}\ \emph {et~al.}(2020)\citenamefont {Linardatos}, \citenamefont {Papastefanopoulos},\ and\ \citenamefont {Kotsiantis}}]{linardatos2020explainable}%
  \BibitemOpen
  \bibfield  {author} {\bibinfo {author} {\bibfnamefont {P.}~\bibnamefont {Linardatos}}, \bibinfo {author} {\bibfnamefont {V.}~\bibnamefont {Papastefanopoulos}},\ and\ \bibinfo {author} {\bibfnamefont {S.}~\bibnamefont {Kotsiantis}},\ }\bibfield  {title} {\bibinfo {title} {Explainable ai: A review of machine learning interpretability methods},\ }\href@noop {} {\bibfield  {journal} {\bibinfo  {journal} {Entropy}\ }\textbf {\bibinfo {volume} {23}},\ \bibinfo {pages} {18} (\bibinfo {year} {2020})}\BibitemShut {NoStop}%
\bibitem [{\citenamefont {Carvalho}\ \emph {et~al.}(2019)\citenamefont {Carvalho}, \citenamefont {Pereira},\ and\ \citenamefont {Cardoso}}]{carvalho2019machine}%
  \BibitemOpen
  \bibfield  {author} {\bibinfo {author} {\bibfnamefont {D.~V.}\ \bibnamefont {Carvalho}}, \bibinfo {author} {\bibfnamefont {E.~M.}\ \bibnamefont {Pereira}},\ and\ \bibinfo {author} {\bibfnamefont {J.~S.}\ \bibnamefont {Cardoso}},\ }\bibfield  {title} {\bibinfo {title} {Machine learning interpretability: A survey on methods and metrics},\ }\href@noop {} {\bibfield  {journal} {\bibinfo  {journal} {Electronics}\ }\textbf {\bibinfo {volume} {8}},\ \bibinfo {pages} {832} (\bibinfo {year} {2019})}\BibitemShut {NoStop}%
\bibitem [{\citenamefont {Zhang}\ \emph {et~al.}(2021)\citenamefont {Zhang}, \citenamefont {Tiňo}, \citenamefont {Leonardis},\ and\ \citenamefont {Tang}}]{zhang2021asurvey}%
  \BibitemOpen
  \bibfield  {author} {\bibinfo {author} {\bibfnamefont {Y.}~\bibnamefont {Zhang}}, \bibinfo {author} {\bibfnamefont {P.}~\bibnamefont {Tiňo}}, \bibinfo {author} {\bibfnamefont {A.}~\bibnamefont {Leonardis}},\ and\ \bibinfo {author} {\bibfnamefont {K.}~\bibnamefont {Tang}},\ }\bibfield  {title} {\bibinfo {title} {A survey on neural network interpretability},\ }\href {https://doi.org/10.1109/TETCI.2021.3100641} {\bibfield  {journal} {\bibinfo  {journal} {IEEE Transactions on Emerging Topics in Computational Intelligence}\ }\textbf {\bibinfo {volume} {5}},\ \bibinfo {pages} {726} (\bibinfo {year} {2021})}\BibitemShut {NoStop}%
\bibitem [{\citenamefont {Gao}\ and\ \citenamefont {Guan}(2023)}]{lei2023interpretability}%
  \BibitemOpen
  \bibfield  {author} {\bibinfo {author} {\bibfnamefont {L.}~\bibnamefont {Gao}}\ and\ \bibinfo {author} {\bibfnamefont {L.}~\bibnamefont {Guan}},\ }\bibfield  {title} {\bibinfo {title} {Interpretability of machine learning: Recent advances and future prospects},\ }\href {https://doi.org/10.1109/MMUL.2023.3272513} {\bibfield  {journal} {\bibinfo  {journal} {IEEE MultiMedia}\ }\textbf {\bibinfo {volume} {30}},\ \bibinfo {pages} {105} (\bibinfo {year} {2023})}\BibitemShut {NoStop}%
\bibitem [{\citenamefont {{\v{S}}trumbelj}\ and\ \citenamefont {Kononenko}(2014)}]{vstrumbelj2014explaining}%
  \BibitemOpen
  \bibfield  {author} {\bibinfo {author} {\bibfnamefont {E.}~\bibnamefont {{\v{S}}trumbelj}}\ and\ \bibinfo {author} {\bibfnamefont {I.}~\bibnamefont {Kononenko}},\ }\bibfield  {title} {\bibinfo {title} {Explaining prediction models and individual predictions with feature contributions},\ }\href@noop {} {\bibfield  {journal} {\bibinfo  {journal} {Knowledge and information systems}\ }\textbf {\bibinfo {volume} {41}},\ \bibinfo {pages} {647} (\bibinfo {year} {2014})}\BibitemShut {NoStop}%
\bibitem [{\citenamefont {Wachter}\ \emph {et~al.}(2017)\citenamefont {Wachter}, \citenamefont {Mittelstadt},\ and\ \citenamefont {Russell}}]{wachter2017counterfactual}%
  \BibitemOpen
  \bibfield  {author} {\bibinfo {author} {\bibfnamefont {S.}~\bibnamefont {Wachter}}, \bibinfo {author} {\bibfnamefont {B.}~\bibnamefont {Mittelstadt}},\ and\ \bibinfo {author} {\bibfnamefont {C.}~\bibnamefont {Russell}},\ }\bibfield  {title} {\bibinfo {title} {Counterfactual explanations without opening the black box: Automated decisions and the gdpr},\ }\href@noop {} {\bibfield  {journal} {\bibinfo  {journal} {Harv. JL \& Tech.}\ }\textbf {\bibinfo {volume} {31}},\ \bibinfo {pages} {841} (\bibinfo {year} {2017})}\BibitemShut {NoStop}%
\bibitem [{\citenamefont {Simonyan}\ \emph {et~al.}(2013)\citenamefont {Simonyan}, \citenamefont {Vedaldi},\ and\ \citenamefont {Zisserman}}]{simonyan2013deep}%
  \BibitemOpen
  \bibfield  {author} {\bibinfo {author} {\bibfnamefont {K.}~\bibnamefont {Simonyan}}, \bibinfo {author} {\bibfnamefont {A.}~\bibnamefont {Vedaldi}},\ and\ \bibinfo {author} {\bibfnamefont {A.}~\bibnamefont {Zisserman}},\ }\bibfield  {title} {\bibinfo {title} {Deep inside convolutional networks: Visualising image classification models and saliency maps},\ }\href@noop {} {\bibfield  {journal} {\bibinfo  {journal} {arXiv preprint arXiv:1312.6034}\ } (\bibinfo {year} {2013})}\BibitemShut {NoStop}%
\bibitem [{\citenamefont {Minh}\ \emph {et~al.}(2022)\citenamefont {Minh}, \citenamefont {Wang}, \citenamefont {Li},\ and\ \citenamefont {Nguyen}}]{minh2022explainable}%
  \BibitemOpen
  \bibfield  {author} {\bibinfo {author} {\bibfnamefont {D.}~\bibnamefont {Minh}}, \bibinfo {author} {\bibfnamefont {H.~X.}\ \bibnamefont {Wang}}, \bibinfo {author} {\bibfnamefont {Y.~F.}\ \bibnamefont {Li}},\ and\ \bibinfo {author} {\bibfnamefont {T.~N.}\ \bibnamefont {Nguyen}},\ }\bibfield  {title} {\bibinfo {title} {Explainable artificial intelligence: A comprehensive review},\ }\href@noop {} {\bibfield  {journal} {\bibinfo  {journal} {Artificial Intelligence Review}\ ,\ \bibinfo {pages} {1}} (\bibinfo {year} {2022})}\BibitemShut {NoStop}%
\bibitem [{\citenamefont {Saeed}\ and\ \citenamefont {Omlin}(2023)}]{saeed2023explainable}%
  \BibitemOpen
  \bibfield  {author} {\bibinfo {author} {\bibfnamefont {W.}~\bibnamefont {Saeed}}\ and\ \bibinfo {author} {\bibfnamefont {C.}~\bibnamefont {Omlin}},\ }\bibfield  {title} {\bibinfo {title} {Explainable ai (xai): A systematic meta-survey of current challenges and future opportunities},\ }\href {https://doi.org/https://doi.org/10.1016/j.knosys.2023.110273} {\bibfield  {journal} {\bibinfo  {journal} {Knowledge-Based Systems}\ }\textbf {\bibinfo {volume} {263}},\ \bibinfo {pages} {110273} (\bibinfo {year} {2023})}\BibitemShut {NoStop}%
\bibitem [{\citenamefont {Hassija}\ \emph {et~al.}(2024)\citenamefont {Hassija}, \citenamefont {Chamola}, \citenamefont {Mahapatra}, \citenamefont {Singal}, \citenamefont {Goel}, \citenamefont {Huang}, \citenamefont {Scardapane}, \citenamefont {Spinelli}, \citenamefont {Mahmud},\ and\ \citenamefont {Hussain}}]{hassija2024interpreting}%
  \BibitemOpen
  \bibfield  {author} {\bibinfo {author} {\bibfnamefont {V.}~\bibnamefont {Hassija}}, \bibinfo {author} {\bibfnamefont {V.}~\bibnamefont {Chamola}}, \bibinfo {author} {\bibfnamefont {A.}~\bibnamefont {Mahapatra}}, \bibinfo {author} {\bibfnamefont {A.}~\bibnamefont {Singal}}, \bibinfo {author} {\bibfnamefont {D.}~\bibnamefont {Goel}}, \bibinfo {author} {\bibfnamefont {K.}~\bibnamefont {Huang}}, \bibinfo {author} {\bibfnamefont {S.}~\bibnamefont {Scardapane}}, \bibinfo {author} {\bibfnamefont {I.}~\bibnamefont {Spinelli}}, \bibinfo {author} {\bibfnamefont {M.}~\bibnamefont {Mahmud}},\ and\ \bibinfo {author} {\bibfnamefont {A.}~\bibnamefont {Hussain}},\ }\bibfield  {title} {\bibinfo {title} {Interpreting black-box models: A review on explainable artificial intelligence},\ }\href@noop {} {\bibfield  {journal} {\bibinfo  {journal} {Cognitive Computation}\ }\textbf {\bibinfo {volume} {16}},\ \bibinfo {pages} {45} (\bibinfo {year} {2024})}\BibitemShut {NoStop}%
\bibitem [{\citenamefont {West}\ \emph {et~al.}(2023{\natexlab{a}})\citenamefont {West}, \citenamefont {Tsang}, \citenamefont {Low}, \citenamefont {Hill}, \citenamefont {Leckie}, \citenamefont {Hollenberg}, \citenamefont {Erfani},\ and\ \citenamefont {Usman}}]{west2023towards}%
  \BibitemOpen
  \bibfield  {author} {\bibinfo {author} {\bibfnamefont {M.~T.}\ \bibnamefont {West}}, \bibinfo {author} {\bibfnamefont {S.-L.}\ \bibnamefont {Tsang}}, \bibinfo {author} {\bibfnamefont {J.~S.}\ \bibnamefont {Low}}, \bibinfo {author} {\bibfnamefont {C.~D.}\ \bibnamefont {Hill}}, \bibinfo {author} {\bibfnamefont {C.}~\bibnamefont {Leckie}}, \bibinfo {author} {\bibfnamefont {L.~C.}\ \bibnamefont {Hollenberg}}, \bibinfo {author} {\bibfnamefont {S.~M.}\ \bibnamefont {Erfani}},\ and\ \bibinfo {author} {\bibfnamefont {M.}~\bibnamefont {Usman}},\ }\bibfield  {title} {\bibinfo {title} {Towards quantum enhanced adversarial robustness in machine learning},\ }\href@noop {} {\bibfield  {journal} {\bibinfo  {journal} {Nature Machine Intelligence}\ }\textbf {\bibinfo {volume} {5}},\ \bibinfo {pages} {581} (\bibinfo {year} {2023}{\natexlab{a}})}\BibitemShut {NoStop}%
\bibitem [{\citenamefont {Wendlinger}\ \emph {et~al.}(2024)\citenamefont {Wendlinger}, \citenamefont {Tscharke},\ and\ \citenamefont {Debus}}]{wendlinger2024comparative}%
  \BibitemOpen
  \bibfield  {author} {\bibinfo {author} {\bibfnamefont {M.}~\bibnamefont {Wendlinger}}, \bibinfo {author} {\bibfnamefont {K.}~\bibnamefont {Tscharke}},\ and\ \bibinfo {author} {\bibfnamefont {P.}~\bibnamefont {Debus}},\ }\href@noop {} {\bibinfo {title} {A comparative analysis of adversarial robustness for quantum and classical machine learning models}} (\bibinfo {year} {2024}),\ \Eprint {https://arxiv.org/abs/2404.16154} {arXiv:2404.16154} \BibitemShut {NoStop}%
\bibitem [{\citenamefont {Lu}\ \emph {et~al.}(2020)\citenamefont {Lu}, \citenamefont {Duan},\ and\ \citenamefont {Deng}}]{lu2020quantumadversarial}%
  \BibitemOpen
  \bibfield  {author} {\bibinfo {author} {\bibfnamefont {S.}~\bibnamefont {Lu}}, \bibinfo {author} {\bibfnamefont {L.-M.}\ \bibnamefont {Duan}},\ and\ \bibinfo {author} {\bibfnamefont {D.-L.}\ \bibnamefont {Deng}},\ }\bibfield  {title} {\bibinfo {title} {Quantum adversarial machine learning},\ }\href {https://doi.org/10.1103/PhysRevResearch.2.033212} {\bibfield  {journal} {\bibinfo  {journal} {Phys. Rev. Res.}\ }\textbf {\bibinfo {volume} {2}},\ \bibinfo {pages} {033212} (\bibinfo {year} {2020})}\BibitemShut {NoStop}%
\bibitem [{\citenamefont {Huang}\ and\ \citenamefont {Zhang}(2023)}]{huang2023enhancing}%
  \BibitemOpen
  \bibfield  {author} {\bibinfo {author} {\bibfnamefont {C.}~\bibnamefont {Huang}}\ and\ \bibinfo {author} {\bibfnamefont {S.}~\bibnamefont {Zhang}},\ }\bibfield  {title} {\bibinfo {title} {Enhancing adversarial robustness of quantum neural networks by adding noise layers},\ }\href@noop {} {\bibfield  {journal} {\bibinfo  {journal} {New Journal of Physics}\ }\textbf {\bibinfo {volume} {25}},\ \bibinfo {pages} {083019} (\bibinfo {year} {2023})}\BibitemShut {NoStop}%
\bibitem [{\citenamefont {Gong}\ \emph {et~al.}(2024)\citenamefont {Gong}, \citenamefont {Yuan}, \citenamefont {Li},\ and\ \citenamefont {Deng}}]{gong2024enhancing}%
  \BibitemOpen
  \bibfield  {author} {\bibinfo {author} {\bibfnamefont {W.}~\bibnamefont {Gong}}, \bibinfo {author} {\bibfnamefont {D.}~\bibnamefont {Yuan}}, \bibinfo {author} {\bibfnamefont {W.}~\bibnamefont {Li}},\ and\ \bibinfo {author} {\bibfnamefont {D.-L.}\ \bibnamefont {Deng}},\ }\bibfield  {title} {\bibinfo {title} {Enhancing quantum adversarial robustness by randomized encodings},\ }\href {https://doi.org/10.1103/PhysRevResearch.6.023020} {\bibfield  {journal} {\bibinfo  {journal} {Phys. Rev. Res.}\ }\textbf {\bibinfo {volume} {6}},\ \bibinfo {pages} {023020} (\bibinfo {year} {2024})}\BibitemShut {NoStop}%
\bibitem [{\citenamefont {Liao}\ \emph {et~al.}(2021)\citenamefont {Liao}, \citenamefont {Convy}, \citenamefont {Huggins},\ and\ \citenamefont {Whaley}}]{liao2021robust}%
  \BibitemOpen
  \bibfield  {author} {\bibinfo {author} {\bibfnamefont {H.}~\bibnamefont {Liao}}, \bibinfo {author} {\bibfnamefont {I.}~\bibnamefont {Convy}}, \bibinfo {author} {\bibfnamefont {W.~J.}\ \bibnamefont {Huggins}},\ and\ \bibinfo {author} {\bibfnamefont {K.~B.}\ \bibnamefont {Whaley}},\ }\bibfield  {title} {\bibinfo {title} {Robust in practice: Adversarial attacks on quantum machine learning},\ }\href {https://doi.org/10.1103/PhysRevA.103.042427} {\bibfield  {journal} {\bibinfo  {journal} {Phys. Rev. A}\ }\textbf {\bibinfo {volume} {103}},\ \bibinfo {pages} {042427} (\bibinfo {year} {2021})}\BibitemShut {NoStop}%
\bibitem [{\citenamefont {Guan}\ \emph {et~al.}(2020)\citenamefont {Guan}, \citenamefont {Fang},\ and\ \citenamefont {Ying}}]{guan2020robustness}%
  \BibitemOpen
  \bibfield  {author} {\bibinfo {author} {\bibfnamefont {J.}~\bibnamefont {Guan}}, \bibinfo {author} {\bibfnamefont {W.}~\bibnamefont {Fang}},\ and\ \bibinfo {author} {\bibfnamefont {M.}~\bibnamefont {Ying}},\ }\bibfield  {title} {\bibinfo {title} {Robustness verification of quantum machine learning},\ }\href@noop {} {\bibfield  {journal} {\bibinfo  {journal} {CoRR}\ } (\bibinfo {year} {2020})}\BibitemShut {NoStop}%
\bibitem [{\citenamefont {Wiebe}\ and\ \citenamefont {Kumar}(2018)}]{wiebe2018hardening}%
  \BibitemOpen
  \bibfield  {author} {\bibinfo {author} {\bibfnamefont {N.}~\bibnamefont {Wiebe}}\ and\ \bibinfo {author} {\bibfnamefont {R.~S.~S.}\ \bibnamefont {Kumar}},\ }\bibfield  {title} {\bibinfo {title} {Hardening quantum machine learning against adversaries},\ }\href@noop {} {\bibfield  {journal} {\bibinfo  {journal} {New Journal of Physics}\ }\textbf {\bibinfo {volume} {20}},\ \bibinfo {pages} {123019} (\bibinfo {year} {2018})}\BibitemShut {NoStop}%
\bibitem [{\citenamefont {Dowling}\ \emph {et~al.}(2024)\citenamefont {Dowling}, \citenamefont {West}, \citenamefont {Southwell}, \citenamefont {Nakhl}, \citenamefont {Sevior}, \citenamefont {Usman},\ and\ \citenamefont {Modi}}]{dowling2024adversarial}%
  \BibitemOpen
  \bibfield  {author} {\bibinfo {author} {\bibfnamefont {N.}~\bibnamefont {Dowling}}, \bibinfo {author} {\bibfnamefont {M.~T.}\ \bibnamefont {West}}, \bibinfo {author} {\bibfnamefont {A.}~\bibnamefont {Southwell}}, \bibinfo {author} {\bibfnamefont {A.~C.}\ \bibnamefont {Nakhl}}, \bibinfo {author} {\bibfnamefont {M.}~\bibnamefont {Sevior}}, \bibinfo {author} {\bibfnamefont {M.}~\bibnamefont {Usman}},\ and\ \bibinfo {author} {\bibfnamefont {K.}~\bibnamefont {Modi}},\ }\href@noop {} {\bibinfo {title} {Adversarial robustness guarantees for quantum classifiers}} (\bibinfo {year} {2024}),\ \Eprint {https://arxiv.org/abs/2405.10360} {arXiv:2405.10360} \BibitemShut {NoStop}%
\bibitem [{\citenamefont {West}\ \emph {et~al.}(2023{\natexlab{b}})\citenamefont {West}, \citenamefont {Erfani}, \citenamefont {Leckie}, \citenamefont {Sevior}, \citenamefont {Hollenberg},\ and\ \citenamefont {Usman}}]{west2023benchmarking}%
  \BibitemOpen
  \bibfield  {author} {\bibinfo {author} {\bibfnamefont {M.~T.}\ \bibnamefont {West}}, \bibinfo {author} {\bibfnamefont {S.~M.}\ \bibnamefont {Erfani}}, \bibinfo {author} {\bibfnamefont {C.}~\bibnamefont {Leckie}}, \bibinfo {author} {\bibfnamefont {M.}~\bibnamefont {Sevior}}, \bibinfo {author} {\bibfnamefont {L.~C.~L.}\ \bibnamefont {Hollenberg}},\ and\ \bibinfo {author} {\bibfnamefont {M.}~\bibnamefont {Usman}},\ }\bibfield  {title} {\bibinfo {title} {Benchmarking adversarially robust quantum machine learning at scale},\ }\href {https://doi.org/10.1103/PhysRevResearch.5.023186} {\bibfield  {journal} {\bibinfo  {journal} {Phys. Rev. Res.}\ }\textbf {\bibinfo {volume} {5}},\ \bibinfo {pages} {023186} (\bibinfo {year} {2023}{\natexlab{b}})}\BibitemShut {NoStop}%
\bibitem [{\citenamefont {Winderl}\ \emph {et~al.}(2024)\citenamefont {Winderl}, \citenamefont {Franco},\ and\ \citenamefont {Lorenz}}]{winderl2024constructing}%
  \BibitemOpen
  \bibfield  {author} {\bibinfo {author} {\bibfnamefont {D.}~\bibnamefont {Winderl}}, \bibinfo {author} {\bibfnamefont {N.}~\bibnamefont {Franco}},\ and\ \bibinfo {author} {\bibfnamefont {J.~M.}\ \bibnamefont {Lorenz}},\ }\bibfield  {title} {\bibinfo {title} {Constructing optimal noise channels for enhanced robustness in quantum machine learning},\ }\href@noop {} {\bibfield  {journal} {\bibinfo  {journal} {arXiv preprint arXiv:2404.16417}\ } (\bibinfo {year} {2024})}\BibitemShut {NoStop}%
\bibitem [{\citenamefont {Sahdev}\ and\ \citenamefont {Kumar}(2023)}]{sahdev2023adversarial}%
  \BibitemOpen
  \bibfield  {author} {\bibinfo {author} {\bibfnamefont {A.}~\bibnamefont {Sahdev}}\ and\ \bibinfo {author} {\bibfnamefont {M.}~\bibnamefont {Kumar}},\ }\href {https://openreview.net/forum?id=o-Yxq5iicIp} {\bibinfo {title} {Adversarial robustness based on randomized smoothing in quantum machine learning}} (\bibinfo {year} {2023})\BibitemShut {NoStop}%
\bibitem [{\citenamefont {Huang}\ \emph {et~al.}(2023)\citenamefont {Huang}, \citenamefont {Tsai}, \citenamefont {Yang}, \citenamefont {Su}, \citenamefont {Yu}, \citenamefont {Chen},\ and\ \citenamefont {Kuo}}]{huang2023certified}%
  \BibitemOpen
  \bibfield  {author} {\bibinfo {author} {\bibfnamefont {J.-C.}\ \bibnamefont {Huang}}, \bibinfo {author} {\bibfnamefont {Y.-L.}\ \bibnamefont {Tsai}}, \bibinfo {author} {\bibfnamefont {C.-H.~H.}\ \bibnamefont {Yang}}, \bibinfo {author} {\bibfnamefont {C.-F.}\ \bibnamefont {Su}}, \bibinfo {author} {\bibfnamefont {C.-M.}\ \bibnamefont {Yu}}, \bibinfo {author} {\bibfnamefont {P.-Y.}\ \bibnamefont {Chen}},\ and\ \bibinfo {author} {\bibfnamefont {S.-Y.}\ \bibnamefont {Kuo}},\ }\bibfield  {title} {\bibinfo {title} {Certified robustness of quantum classifiers against adversarial examples through quantum noise},\ }in\ \href {https://doi.org/10.1109/ICASSP49357.2023.10095030} {\emph {\bibinfo {booktitle} {ICASSP 2023 - 2023 IEEE International Conference on Acoustics, Speech and Signal Processing (ICASSP)}}}\ (\bibinfo {year} {2023})\ pp.\ \bibinfo {pages} {1--5}\BibitemShut {NoStop}%
\bibitem [{\citenamefont {Du}\ \emph {et~al.}(2021)\citenamefont {Du}, \citenamefont {Hsieh}, \citenamefont {Liu}, \citenamefont {Tao},\ and\ \citenamefont {Liu}}]{du2021quantumnoise}%
  \BibitemOpen
  \bibfield  {author} {\bibinfo {author} {\bibfnamefont {Y.}~\bibnamefont {Du}}, \bibinfo {author} {\bibfnamefont {M.-H.}\ \bibnamefont {Hsieh}}, \bibinfo {author} {\bibfnamefont {T.}~\bibnamefont {Liu}}, \bibinfo {author} {\bibfnamefont {D.}~\bibnamefont {Tao}},\ and\ \bibinfo {author} {\bibfnamefont {N.}~\bibnamefont {Liu}},\ }\bibfield  {title} {\bibinfo {title} {Quantum noise protects quantum classifiers against adversaries},\ }\href {https://doi.org/10.1103/PhysRevResearch.3.023153} {\bibfield  {journal} {\bibinfo  {journal} {Phys. Rev. Res.}\ }\textbf {\bibinfo {volume} {3}},\ \bibinfo {pages} {023153} (\bibinfo {year} {2021})}\BibitemShut {NoStop}%
\bibitem [{\citenamefont {Franco}\ \emph {et~al.}(2024)\citenamefont {Franco}, \citenamefont {Sakhnenko}, \citenamefont {Stolpmann}, \citenamefont {Thuerck}, \citenamefont {Petsch}, \citenamefont {R{\"u}ll},\ and\ \citenamefont {Lorenz}}]{franco2024predominant}%
  \BibitemOpen
  \bibfield  {author} {\bibinfo {author} {\bibfnamefont {N.}~\bibnamefont {Franco}}, \bibinfo {author} {\bibfnamefont {A.}~\bibnamefont {Sakhnenko}}, \bibinfo {author} {\bibfnamefont {L.}~\bibnamefont {Stolpmann}}, \bibinfo {author} {\bibfnamefont {D.}~\bibnamefont {Thuerck}}, \bibinfo {author} {\bibfnamefont {F.}~\bibnamefont {Petsch}}, \bibinfo {author} {\bibfnamefont {A.}~\bibnamefont {R{\"u}ll}},\ and\ \bibinfo {author} {\bibfnamefont {J.~M.}\ \bibnamefont {Lorenz}},\ }\bibfield  {title} {\bibinfo {title} {Predominant aspects on security for quantum machine learning: Literature review},\ }\href@noop {} {\bibfield  {journal} {\bibinfo  {journal} {arXiv preprint arXiv:2401.07774}\ } (\bibinfo {year} {2024})}\BibitemShut {NoStop}%
\bibitem [{\citenamefont {Berberich}\ \emph {et~al.}(2023)\citenamefont {Berberich}, \citenamefont {Fink}, \citenamefont {Pranji{\'c}}, \citenamefont {Tutschku},\ and\ \citenamefont {Holm}}]{berberich2023training}%
  \BibitemOpen
  \bibfield  {author} {\bibinfo {author} {\bibfnamefont {J.}~\bibnamefont {Berberich}}, \bibinfo {author} {\bibfnamefont {D.}~\bibnamefont {Fink}}, \bibinfo {author} {\bibfnamefont {D.}~\bibnamefont {Pranji{\'c}}}, \bibinfo {author} {\bibfnamefont {C.}~\bibnamefont {Tutschku}},\ and\ \bibinfo {author} {\bibfnamefont {C.}~\bibnamefont {Holm}},\ }\bibfield  {title} {\bibinfo {title} {Training robust and generalizable quantum models},\ }\href@noop {} {\bibfield  {journal} {\bibinfo  {journal} {arXiv preprint arXiv:2311.11871}\ } (\bibinfo {year} {2023})}\BibitemShut {NoStop}%
\bibitem [{\citenamefont {Guan}\ \emph {et~al.}(2021)\citenamefont {Guan}, \citenamefont {Fang},\ and\ \citenamefont {Ying}}]{guan2021robustness}%
  \BibitemOpen
  \bibfield  {author} {\bibinfo {author} {\bibfnamefont {J.}~\bibnamefont {Guan}}, \bibinfo {author} {\bibfnamefont {W.}~\bibnamefont {Fang}},\ and\ \bibinfo {author} {\bibfnamefont {M.}~\bibnamefont {Ying}},\ }\bibfield  {title} {\bibinfo {title} {Robustness verification of quantum classifiers},\ }in\ \href@noop {} {\emph {\bibinfo {booktitle} {Computer Aided Verification: 33rd International Conference, CAV 2021, Virtual Event, July 20--23, 2021, Proceedings, Part I 33}}}\ (\bibinfo {organization} {Springer},\ \bibinfo {year} {2021})\ pp.\ \bibinfo {pages} {151--174}\BibitemShut {NoStop}%
\bibitem [{\citenamefont {Weber}\ \emph {et~al.}(2021)\citenamefont {Weber}, \citenamefont {Liu}, \citenamefont {Li}, \citenamefont {Zhang},\ and\ \citenamefont {Zhao}}]{weber2021optimal}%
  \BibitemOpen
  \bibfield  {author} {\bibinfo {author} {\bibfnamefont {M.}~\bibnamefont {Weber}}, \bibinfo {author} {\bibfnamefont {N.}~\bibnamefont {Liu}}, \bibinfo {author} {\bibfnamefont {B.}~\bibnamefont {Li}}, \bibinfo {author} {\bibfnamefont {C.}~\bibnamefont {Zhang}},\ and\ \bibinfo {author} {\bibfnamefont {Z.}~\bibnamefont {Zhao}},\ }\bibfield  {title} {\bibinfo {title} {Optimal provable robustness of quantum classification via quantum hypothesis testing},\ }\href@noop {} {\bibfield  {journal} {\bibinfo  {journal} {npj Quantum Information}\ }\textbf {\bibinfo {volume} {7}},\ \bibinfo {pages} {76} (\bibinfo {year} {2021})}\BibitemShut {NoStop}%
\bibitem [{\citenamefont {Steinm{\"u}ller}\ \emph {et~al.}(2022)\citenamefont {Steinm{\"u}ller}, \citenamefont {Schulz}, \citenamefont {Graf},\ and\ \citenamefont {Herr}}]{steinmuller2022explainable}%
  \BibitemOpen
  \bibfield  {author} {\bibinfo {author} {\bibfnamefont {P.}~\bibnamefont {Steinm{\"u}ller}}, \bibinfo {author} {\bibfnamefont {T.}~\bibnamefont {Schulz}}, \bibinfo {author} {\bibfnamefont {F.}~\bibnamefont {Graf}},\ and\ \bibinfo {author} {\bibfnamefont {D.}~\bibnamefont {Herr}},\ }\bibfield  {title} {\bibinfo {title} {Explainable ai for quantum machine learning},\ }\href@noop {} {\bibfield  {journal} {\bibinfo  {journal} {arXiv preprint arXiv:2211.01441}\ } (\bibinfo {year} {2022})}\BibitemShut {NoStop}%
\bibitem [{\citenamefont {Power}\ and\ \citenamefont {Guha}(2024)}]{power2024feature}%
  \BibitemOpen
  \bibfield  {author} {\bibinfo {author} {\bibfnamefont {L.}~\bibnamefont {Power}}\ and\ \bibinfo {author} {\bibfnamefont {K.}~\bibnamefont {Guha}},\ }\href@noop {} {\bibinfo {title} {Feature importance and explainability in quantum machine learning}} (\bibinfo {year} {2024}),\ \Eprint {https://arxiv.org/abs/2405.08917} {arXiv:2405.08917} \BibitemShut {NoStop}%
\bibitem [{\citenamefont {Liu}\ \emph {et~al.}(2022)\citenamefont {Liu}, \citenamefont {Shen}, \citenamefont {Li}, \citenamefont {Duan},\ and\ \citenamefont {Deng}}]{liu2023quantum}%
  \BibitemOpen
  \bibfield  {author} {\bibinfo {author} {\bibfnamefont {Z.}~\bibnamefont {Liu}}, \bibinfo {author} {\bibfnamefont {P.-X.}\ \bibnamefont {Shen}}, \bibinfo {author} {\bibfnamefont {W.}~\bibnamefont {Li}}, \bibinfo {author} {\bibfnamefont {L.-M.}\ \bibnamefont {Duan}},\ and\ \bibinfo {author} {\bibfnamefont {D.-L.}\ \bibnamefont {Deng}},\ }\bibfield  {title} {\bibinfo {title} {Quantum capsule networks},\ }\href {https://doi.org/10.1088/2058-9565/aca55d} {\bibfield  {journal} {\bibinfo  {journal} {Quantum Science and Technology}\ }\textbf {\bibinfo {volume} {8}},\ \bibinfo {pages} {015016} (\bibinfo {year} {2022})}\BibitemShut {NoStop}%
\bibitem [{\citenamefont {Heese}\ \emph {et~al.}(2023)\citenamefont {Heese}, \citenamefont {Gerlach}, \citenamefont {Mücke}, \citenamefont {Müller}, \citenamefont {Jakobs},\ and\ \citenamefont {Piatkowski}}]{heese2023explaining}%
  \BibitemOpen
  \bibfield  {author} {\bibinfo {author} {\bibfnamefont {R.}~\bibnamefont {Heese}}, \bibinfo {author} {\bibfnamefont {T.}~\bibnamefont {Gerlach}}, \bibinfo {author} {\bibfnamefont {S.}~\bibnamefont {Mücke}}, \bibinfo {author} {\bibfnamefont {S.}~\bibnamefont {Müller}}, \bibinfo {author} {\bibfnamefont {M.}~\bibnamefont {Jakobs}},\ and\ \bibinfo {author} {\bibfnamefont {N.}~\bibnamefont {Piatkowski}},\ }\href@noop {} {\bibinfo {title} {Explaining quantum circuits with shapley values: Towards explainable quantum machine learning}} (\bibinfo {year} {2023}),\ \Eprint {https://arxiv.org/abs/2301.09138} {arXiv:2301.09138} \BibitemShut {NoStop}%
\bibitem [{\citenamefont {Ruan}\ \emph {et~al.}(2024)\citenamefont {Ruan}, \citenamefont {Liang}, \citenamefont {Guan}, \citenamefont {Griffin}, \citenamefont {Wen}, \citenamefont {Lin},\ and\ \citenamefont {Wang}}]{shaolun2024violet}%
  \BibitemOpen
  \bibfield  {author} {\bibinfo {author} {\bibfnamefont {S.}~\bibnamefont {Ruan}}, \bibinfo {author} {\bibfnamefont {Z.}~\bibnamefont {Liang}}, \bibinfo {author} {\bibfnamefont {Q.}~\bibnamefont {Guan}}, \bibinfo {author} {\bibfnamefont {P.}~\bibnamefont {Griffin}}, \bibinfo {author} {\bibfnamefont {X.}~\bibnamefont {Wen}}, \bibinfo {author} {\bibfnamefont {Y.}~\bibnamefont {Lin}},\ and\ \bibinfo {author} {\bibfnamefont {Y.}~\bibnamefont {Wang}},\ }\bibfield  {title} {\bibinfo {title} {Violet: Visual analytics for explainable quantum neural networks},\ }\href {https://doi.org/10.1109/TVCG.2024.3388557} {\bibfield  {journal} {\bibinfo  {journal} {IEEE Transactions on Visualization and Computer Graphics}\ }\textbf {\bibinfo {volume} {30}},\ \bibinfo {pages} {2862} (\bibinfo {year} {2024})}\BibitemShut {NoStop}%
\bibitem [{\citenamefont {Baughman}\ \emph {et~al.}(2022)\citenamefont {Baughman}, \citenamefont {Yogaraj}, \citenamefont {Hebbar}, \citenamefont {Ghosh}, \citenamefont {Haq},\ and\ \citenamefont {Chhabra}}]{baughman2022study}%
  \BibitemOpen
  \bibfield  {author} {\bibinfo {author} {\bibfnamefont {A.}~\bibnamefont {Baughman}}, \bibinfo {author} {\bibfnamefont {K.}~\bibnamefont {Yogaraj}}, \bibinfo {author} {\bibfnamefont {R.}~\bibnamefont {Hebbar}}, \bibinfo {author} {\bibfnamefont {S.}~\bibnamefont {Ghosh}}, \bibinfo {author} {\bibfnamefont {R.~U.}\ \bibnamefont {Haq}},\ and\ \bibinfo {author} {\bibfnamefont {Y.}~\bibnamefont {Chhabra}},\ }\href@noop {} {\bibinfo {title} {Study of feature importance for quantum machine learning models}} (\bibinfo {year} {2022}),\ \Eprint {https://arxiv.org/abs/2202.11204} {arXiv:2202.11204} \BibitemShut {NoStop}%
\bibitem [{\citenamefont {Gil-Fuster}\ \emph {et~al.}(2024{\natexlab{a}})\citenamefont {Gil-Fuster}, \citenamefont {Naujoks}, \citenamefont {Montavon}, \citenamefont {Wiegand}, \citenamefont {Samek},\ and\ \citenamefont {Eisert}}]{gil2024opportunities}%
  \BibitemOpen
  \bibfield  {author} {\bibinfo {author} {\bibfnamefont {E.}~\bibnamefont {Gil-Fuster}}, \bibinfo {author} {\bibfnamefont {J.~R.}\ \bibnamefont {Naujoks}}, \bibinfo {author} {\bibfnamefont {G.}~\bibnamefont {Montavon}}, \bibinfo {author} {\bibfnamefont {T.}~\bibnamefont {Wiegand}}, \bibinfo {author} {\bibfnamefont {W.}~\bibnamefont {Samek}},\ and\ \bibinfo {author} {\bibfnamefont {J.}~\bibnamefont {Eisert}},\ }\bibfield  {title} {\bibinfo {title} {Opportunities and limitations of explaining quantum machine learning},\ }\href@noop {} {\bibfield  {journal} {\bibinfo  {journal} {arXiv preprint arXiv:2412.14753}\ } (\bibinfo {year} {2024}{\natexlab{a}})}\BibitemShut {NoStop}%
\bibitem [{\citenamefont {Anschuetz}\ \emph {et~al.}(2023)\citenamefont {Anschuetz}, \citenamefont {Hu}, \citenamefont {Huang},\ and\ \citenamefont {Gao}}]{anschuetz2023interpretable}%
  \BibitemOpen
  \bibfield  {author} {\bibinfo {author} {\bibfnamefont {E.~R.}\ \bibnamefont {Anschuetz}}, \bibinfo {author} {\bibfnamefont {H.-Y.}\ \bibnamefont {Hu}}, \bibinfo {author} {\bibfnamefont {J.-L.}\ \bibnamefont {Huang}},\ and\ \bibinfo {author} {\bibfnamefont {X.}~\bibnamefont {Gao}},\ }\bibfield  {title} {\bibinfo {title} {Interpretable quantum advantage in neural sequence learning},\ }\href {https://doi.org/10.1103/PRXQuantum.4.020338} {\bibfield  {journal} {\bibinfo  {journal} {PRX Quantum}\ }\textbf {\bibinfo {volume} {4}},\ \bibinfo {pages} {020338} (\bibinfo {year} {2023})}\BibitemShut {NoStop}%
\bibitem [{\citenamefont {Ruan}\ \emph {et~al.}(2023)\citenamefont {Ruan}, \citenamefont {Guan}, \citenamefont {Griffin}, \citenamefont {Mao},\ and\ \citenamefont {Wang}}]{shaolun2023quantumeyes}%
  \BibitemOpen
  \bibfield  {author} {\bibinfo {author} {\bibfnamefont {S.}~\bibnamefont {Ruan}}, \bibinfo {author} {\bibfnamefont {Q.}~\bibnamefont {Guan}}, \bibinfo {author} {\bibfnamefont {P.}~\bibnamefont {Griffin}}, \bibinfo {author} {\bibfnamefont {Y.}~\bibnamefont {Mao}},\ and\ \bibinfo {author} {\bibfnamefont {Y.}~\bibnamefont {Wang}},\ }\bibfield  {title} {\bibinfo {title} {Quantumeyes: Towards better interpretability of quantum circuits},\ }\href {https://doi.org/10.1109/TVCG.2023.3332999} {\bibfield  {journal} {\bibinfo  {journal} {IEEE Transactions on Visualization and Computer Graphics}\ ,\ \bibinfo {pages} {1}} (\bibinfo {year} {2023})}\BibitemShut {NoStop}%
\bibitem [{\citenamefont {Perrier}(2021)}]{perrier2021quantumfair}%
  \BibitemOpen
  \bibfield  {author} {\bibinfo {author} {\bibfnamefont {E.}~\bibnamefont {Perrier}},\ }\bibfield  {title} {\bibinfo {title} {Quantum fair machine learning},\ }in\ \href {https://doi.org/10.1145/3461702.3462611} {\emph {\bibinfo {booktitle} {Proceedings of the 2021 AAAI/ACM Conference on AI, Ethics, and Society}}},\ \bibinfo {series and number} {AIES '21}\ (\bibinfo  {publisher} {Association for Computing Machinery},\ \bibinfo {address} {New York, NY, USA},\ \bibinfo {year} {2021})\ p.\ \bibinfo {pages} {843–853}\BibitemShut {NoStop}%
\bibitem [{\citenamefont {Guan}\ \emph {et~al.}(2022)\citenamefont {Guan}, \citenamefont {Fang},\ and\ \citenamefont {Ying}}]{guan2022verifying}%
  \BibitemOpen
  \bibfield  {author} {\bibinfo {author} {\bibfnamefont {J.}~\bibnamefont {Guan}}, \bibinfo {author} {\bibfnamefont {W.}~\bibnamefont {Fang}},\ and\ \bibinfo {author} {\bibfnamefont {M.}~\bibnamefont {Ying}},\ }\bibfield  {title} {\bibinfo {title} {Verifying fairness in quantum machine learning},\ }in\ \href@noop {} {\emph {\bibinfo {booktitle} {International Conference on Computer Aided Verification}}}\ (\bibinfo {organization} {Springer},\ \bibinfo {year} {2022})\ pp.\ \bibinfo {pages} {408--429}\BibitemShut {NoStop}%
\bibitem [{\citenamefont {Franco}\ \emph {et~al.}(2022)\citenamefont {Franco}, \citenamefont {Wollschl{\"a}ger}, \citenamefont {Gao}, \citenamefont {Lorenz},\ and\ \citenamefont {G{\"u}nnemann}}]{franco2022quantum}%
  \BibitemOpen
  \bibfield  {author} {\bibinfo {author} {\bibfnamefont {N.}~\bibnamefont {Franco}}, \bibinfo {author} {\bibfnamefont {T.}~\bibnamefont {Wollschl{\"a}ger}}, \bibinfo {author} {\bibfnamefont {N.}~\bibnamefont {Gao}}, \bibinfo {author} {\bibfnamefont {J.~M.}\ \bibnamefont {Lorenz}},\ and\ \bibinfo {author} {\bibfnamefont {S.}~\bibnamefont {G{\"u}nnemann}},\ }\bibfield  {title} {\bibinfo {title} {Quantum robustness verification: A hybrid quantum-classical neural network certification algorithm},\ }in\ \href@noop {} {\emph {\bibinfo {booktitle} {2022 IEEE International Conference on Quantum Computing and Engineering (QCE)}}}\ (\bibinfo {organization} {IEEE},\ \bibinfo {year} {2022})\ pp.\ \bibinfo {pages} {142--153}\BibitemShut {NoStop}%
\bibitem [{\citenamefont {Park}\ and\ \citenamefont {Simeone}(2024)}]{park2024quantumCP}%
  \BibitemOpen
  \bibfield  {author} {\bibinfo {author} {\bibfnamefont {S.}~\bibnamefont {Park}}\ and\ \bibinfo {author} {\bibfnamefont {O.}~\bibnamefont {Simeone}},\ }\bibfield  {title} {\bibinfo {title} {Quantum conformal prediction for reliable uncertainty quantification in quantum machine learning},\ }\href {https://doi.org/10.1109/TQE.2023.3333224} {\bibfield  {journal} {\bibinfo  {journal} {IEEE Transactions on Quantum Engineering}\ }\textbf {\bibinfo {volume} {5}},\ \bibinfo {pages} {1} (\bibinfo {year} {2024})}\BibitemShut {NoStop}%
\bibitem [{\citenamefont {Baldassi}\ \emph {et~al.}(2020)\citenamefont {Baldassi}, \citenamefont {Pittorino},\ and\ \citenamefont {Zecchina}}]{baldassi2020shaping}%
  \BibitemOpen
  \bibfield  {author} {\bibinfo {author} {\bibfnamefont {C.}~\bibnamefont {Baldassi}}, \bibinfo {author} {\bibfnamefont {F.}~\bibnamefont {Pittorino}},\ and\ \bibinfo {author} {\bibfnamefont {R.}~\bibnamefont {Zecchina}},\ }\bibfield  {title} {\bibinfo {title} {Shaping the learning landscape in neural networks around wide flat minima},\ }\href@noop {} {\bibfield  {journal} {\bibinfo  {journal} {Proceedings of the National Academy of Sciences}\ }\textbf {\bibinfo {volume} {117}},\ \bibinfo {pages} {161} (\bibinfo {year} {2020})}\BibitemShut {NoStop}%
\bibitem [{\citenamefont {Huembeli}\ and\ \citenamefont {Dauphin}(2021)}]{huembeli2021}%
  \BibitemOpen
  \bibfield  {author} {\bibinfo {author} {\bibfnamefont {P.}~\bibnamefont {Huembeli}}\ and\ \bibinfo {author} {\bibfnamefont {A.}~\bibnamefont {Dauphin}},\ }\bibfield  {title} {\bibinfo {title} {Characterizing the loss landscape of variational quantum circuits},\ }\href {https://doi.org/10.1088/2058-9565/abdbc9} {\bibfield  {journal} {\bibinfo  {journal} {Quantum Science and Technology}\ }\textbf {\bibinfo {volume} {6}},\ \bibinfo {pages} {025011} (\bibinfo {year} {2021})}\BibitemShut {NoStop}%
\bibitem [{\citenamefont {Fisher}(1988)}]{iris}%
  \BibitemOpen
  \bibfield  {author} {\bibinfo {author} {\bibfnamefont {R.~A.}\ \bibnamefont {Fisher}},\ }\href@noop {} {\bibinfo {title} {Iris}},\ \bibinfo {howpublished} {{UCI Machine Learning Repository}} (\bibinfo {year} {1988}),\ \bibinfo {note} {{DOI}: https://doi.org/10.24432/C56C76}\BibitemShut {NoStop}%
\bibitem [{\citenamefont {P{\'e}rez-Salinas}\ \emph {et~al.}(2020)\citenamefont {P{\'e}rez-Salinas}, \citenamefont {Cervera-Lierta}, \citenamefont {Gil-Fuster},\ and\ \citenamefont {Latorre}}]{perez2020data}%
  \BibitemOpen
  \bibfield  {author} {\bibinfo {author} {\bibfnamefont {A.}~\bibnamefont {P{\'e}rez-Salinas}}, \bibinfo {author} {\bibfnamefont {A.}~\bibnamefont {Cervera-Lierta}}, \bibinfo {author} {\bibfnamefont {E.}~\bibnamefont {Gil-Fuster}},\ and\ \bibinfo {author} {\bibfnamefont {J.~I.}\ \bibnamefont {Latorre}},\ }\bibfield  {title} {\bibinfo {title} {Data re-uploading for a universal quantum classifier},\ }\href@noop {} {\bibfield  {journal} {\bibinfo  {journal} {Quantum}\ }\textbf {\bibinfo {volume} {4}},\ \bibinfo {pages} {226} (\bibinfo {year} {2020})}\BibitemShut {NoStop}%
\bibitem [{\citenamefont {Gil-Fuster}\ \emph {et~al.}(2024{\natexlab{b}})\citenamefont {Gil-Fuster}, \citenamefont {Gyurik}, \citenamefont {P{\'e}rez-Salinas},\ and\ \citenamefont {Dunjko}}]{gil2024relation}%
  \BibitemOpen
  \bibfield  {author} {\bibinfo {author} {\bibfnamefont {E.}~\bibnamefont {Gil-Fuster}}, \bibinfo {author} {\bibfnamefont {C.}~\bibnamefont {Gyurik}}, \bibinfo {author} {\bibfnamefont {A.}~\bibnamefont {P{\'e}rez-Salinas}},\ and\ \bibinfo {author} {\bibfnamefont {V.}~\bibnamefont {Dunjko}},\ }\bibfield  {title} {\bibinfo {title} {On the relation between trainability and dequantization of variational quantum learning models},\ }\href@noop {} {\bibfield  {journal} {\bibinfo  {journal} {arXiv preprint arXiv:2406.07072}\ } (\bibinfo {year} {2024}{\natexlab{b}})}\BibitemShut {NoStop}%
\bibitem [{\citenamefont {Thabet}\ \emph {et~al.}(2024)\citenamefont {Thabet}, \citenamefont {Monbroussou}, \citenamefont {Mamon},\ and\ \citenamefont {Landman}}]{thabet2024quantum}%
  \BibitemOpen
  \bibfield  {author} {\bibinfo {author} {\bibfnamefont {S.}~\bibnamefont {Thabet}}, \bibinfo {author} {\bibfnamefont {L.}~\bibnamefont {Monbroussou}}, \bibinfo {author} {\bibfnamefont {E.~Z.}\ \bibnamefont {Mamon}},\ and\ \bibinfo {author} {\bibfnamefont {J.}~\bibnamefont {Landman}},\ }\bibfield  {title} {\bibinfo {title} {When quantum and classical models disagree: Learning beyond minimum norm least square},\ }\href@noop {} {\bibfield  {journal} {\bibinfo  {journal} {arXiv preprint arXiv:2411.04940}\ } (\bibinfo {year} {2024})}\BibitemShut {NoStop}%
\bibitem [{\citenamefont {Liu}\ \emph {et~al.}(2021{\natexlab{d}})\citenamefont {Liu}, \citenamefont {Arunachalam},\ and\ \citenamefont {Temme}}]{liu2021rigorous}%
  \BibitemOpen
  \bibfield  {author} {\bibinfo {author} {\bibfnamefont {Y.}~\bibnamefont {Liu}}, \bibinfo {author} {\bibfnamefont {S.}~\bibnamefont {Arunachalam}},\ and\ \bibinfo {author} {\bibfnamefont {K.}~\bibnamefont {Temme}},\ }\bibfield  {title} {\bibinfo {title} {A rigorous and robust quantum speed-up in supervised machine learning},\ }\href@noop {} {\bibfield  {journal} {\bibinfo  {journal} {Nature Physics}\ }\textbf {\bibinfo {volume} {17}},\ \bibinfo {pages} {1013} (\bibinfo {year} {2021}{\natexlab{d}})}\BibitemShut {NoStop}%
\bibitem [{\citenamefont {Schuld}\ \emph {et~al.}(2021)\citenamefont {Schuld}, \citenamefont {Sweke},\ and\ \citenamefont {Meyer}}]{schuld2021effectofdataencoding}%
  \BibitemOpen
  \bibfield  {author} {\bibinfo {author} {\bibfnamefont {M.}~\bibnamefont {Schuld}}, \bibinfo {author} {\bibfnamefont {R.}~\bibnamefont {Sweke}},\ and\ \bibinfo {author} {\bibfnamefont {J.~J.}\ \bibnamefont {Meyer}},\ }\bibfield  {title} {\bibinfo {title} {Effect of data encoding on the expressive power of variational quantum-machine-learning models},\ }\href {https://doi.org/10.1103/PhysRevA.103.032430} {\bibfield  {journal} {\bibinfo  {journal} {Phys. Rev. A}\ }\textbf {\bibinfo {volume} {103}},\ \bibinfo {pages} {032430} (\bibinfo {year} {2021})}\BibitemShut {NoStop}%
\bibitem [{\citenamefont {Pedregosa}\ \emph {et~al.}(2011)\citenamefont {Pedregosa}, \citenamefont {Varoquaux}, \citenamefont {Gramfort}, \citenamefont {Michel}, \citenamefont {Thirion}, \citenamefont {Grisel}, \citenamefont {Blondel}, \citenamefont {Prettenhofer}, \citenamefont {Weiss}, \citenamefont {Dubourg} \emph {et~al.}}]{pedregosa2011scikit}%
  \BibitemOpen
  \bibfield  {author} {\bibinfo {author} {\bibfnamefont {F.}~\bibnamefont {Pedregosa}}, \bibinfo {author} {\bibfnamefont {G.}~\bibnamefont {Varoquaux}}, \bibinfo {author} {\bibfnamefont {A.}~\bibnamefont {Gramfort}}, \bibinfo {author} {\bibfnamefont {V.}~\bibnamefont {Michel}}, \bibinfo {author} {\bibfnamefont {B.}~\bibnamefont {Thirion}}, \bibinfo {author} {\bibfnamefont {O.}~\bibnamefont {Grisel}}, \bibinfo {author} {\bibfnamefont {M.}~\bibnamefont {Blondel}}, \bibinfo {author} {\bibfnamefont {P.}~\bibnamefont {Prettenhofer}}, \bibinfo {author} {\bibfnamefont {R.}~\bibnamefont {Weiss}}, \bibinfo {author} {\bibfnamefont {V.}~\bibnamefont {Dubourg}}, \emph {et~al.},\ }\bibfield  {title} {\bibinfo {title} {Scikit-learn: Machine learning in python},\ }\href {https://dl.acm.org/doi/10.5555/1953048.2078195} {\bibfield  {journal} {\bibinfo  {journal} {the Journal of Machine Learning research}\ }\textbf {\bibinfo {volume} {12}},\ \bibinfo {pages} {2825} (\bibinfo {year} {2011})}\BibitemShut {NoStop}%
\end{thebibliography}%

\end{document}